\documentclass[preprint, amsmath, amssymb, aps, physrev]{revtex4-2}
\usepackage{amsmath}
\usepackage{amsfonts}
\usepackage{hyperref}
\hypersetup{
    colorlinks = true,
    citecolor = blue,
    linkcolor = red,
}
\usepackage{graphicx}
\usepackage{algorithm}
\usepackage{algpseudocode}

\begin{document}

\title{Higher-Order Transverse Discontinuity Mapping in Filippov Systems: Analysis and Experimental Validation using an Electronic Circuit}


\author{Rohit Chawla}
\email{rohit.chawla@ucd.ie}
\author{Soumyajit Seth}
\email{soumyajit.seth@nmims.edu}
\author{Aasifa Rounak}
\email{aasifa.rounak@ucd.ie}
\author{Vikram Pakrashi}
\email{vikram.pakrashi@ucd.ie}

\affiliation{UCD Centre for Mechanics, Dynamical Systems and Risk Laboratory, School of Mechanical and Materials Engineering, University College Dublin, Ireland}
\affiliation{School of Technology Management \& Engineering, SVKM's Narsee Monjee Institute of Management Studies (NMIMS) Deemed-to-University, Hyderabad, 509301, Telangana, India.}

\begin{abstract}
This paper shows that linearizing the transverse discontinuity mapping (TDM) in Filippov systems can produce inaccurate predictions of the dynamics in impact oscillators operating near a pre-stressed soft barrier.
This discrepancy arises from the limitations of the linearized saltation matrix, which inaccurately predicts impacts in the local neighborhood of the discontinuity boundary. To address this issue, a higher-order approximation of the TDM is derived, which accurately captures the onset of impacts and closely matches the results obtained from both numerical simulations and electronic experiments. The proposed higher-order TDM results in a quadratic estimation of flight time for impacts. Geometrically, real-valued impact events are only feasible when the discriminant of this quadratic equation is positive. The differences in the predicted higher-order flight times and mapping estimates become more pronounced for low-velocity impacts close to grazing. Subsequently, a numerical approximation of the higher-order saltation matrix and its consequent Floquet multipliers and Lyapunov exponents for stability analysis is proposed and demonstrated on a pre-stressed, forced, damped, soft impact oscillator. To validate the numerically observed discontinuity-induced bifurcations, an analog electronic circuit is proposed that models a soft mechanical prestressed oscillator. This inductor-less circuit overcomes the limitations of typical LCR-based circuits, which are used to design such oscillators but cannot accommodate low stiffness ratios. The experimentally obtained limit cycles, finger-shaped Poincar\'e sections, and bifurcation diagrams match the predictions of the higher-order TDM, validating that the proposed circuit accurately models the soft-impact oscillator for both low and high stiffness ratios.
\end{abstract}

\maketitle

\section{Introduction}

Discontinuity-induced bifurcations (DIBs) commonly arise in dynamical systems due to frictional interactions and impacts between moving components during operation. Such impacts may result from wear and tear, limited mechanical clearance, or loosened joints. Examples include impacts between print hammers~\cite{hendriks1983bounce,tung1988method}, heat exchanger tubes~\cite{goyder1989study,paidoussis1992cross}, walking robots~\cite{holmes2006dynamics}, floating or buoyant vessels striking rigid harbors during ship grounding events~\cite{virgin2009some,ibrahim2014recent,xue2023nonlinear}, vibro-impacting structures subjected to vortex-induced vibrations~\cite{chawla2024wake}, and many-body systems influenced by frictional contact~\cite{feeny1992nonsmooth,fan2020discontinuous}.

The dynamics of such systems are often studied through rigid~\cite{nordmark1991non} and soft~\cite{jiang2017grazing} impact oscillator models, which are mathematically formulated as piecewise-smooth dynamical systems~\cite{bernardo2008piecewise}. Examples of DIBs in impact oscillators include period-adding cascades~\cite{piiroinen2004chaos,oestreich1996bifurcation,oestreich1997analytical,chawla2024wake}, grazing bifurcations~\cite{banerjee2009invisible,jiang2017grazing,chin1994grazing}, chattering phenomena~\cite{budd1994chattering}, and narrow-band chaos or dangerous border-collision bifurcations~\cite{banerjee2009invisible}, often accompanied by multiple coexisting attractors~\cite{chawla2024wake}.

These DIBs typically lead to large-amplitude oscillations, primarily due to local stretching in the phase space~\cite{nordmark1991non}, and can be detrimental to the performance and reliability of mechanical systems.

Nordmark showed that DIBs occur for low-velocity impacts near grazing due to the square-root singularity \cite{nordmark1991non}. 
Away from the grazing condition—that is, when orbits interact transversely with the discontinuity boundary—the behavior of orbits is characterized using a \textit{transverse discontinuity map} (TDM). Fredriksson and Nordmark~\cite{fredriksson1997bifurcations} introduced a \textit{discontinuity bypass map} for hard-impact oscillators under grazing incidence, motivated by the normal form derivations of Nordmark~\cite{nordmark1991non, nordmark1992effects}. In a subsequent study, Fredriksson and Nordmark~\cite{fredriksson2000normal} derived mappings relevant to transverse interactions.

Dankowicz and Nordmark~\cite{dankowicz2000origin} extended the study of discontinuity mappings to Filippov systems under grazing conditions. Furthermore, di Bernardo \textit{et al.}~\cite{di2001normal} introduced the \textit{zero-time discontinuity mapping} (ZDM) and the \textit{Poincar\'e section discontinuity mapping} (PDM) to analyze grazing orbits in piecewise-smooth (PWS), piecewise-continuous (PWC), and Filippov systems. Later, Leine~\cite{leine2000bifurcations} derived the \textit{saltation matrix} to describe transverse interactions in Filippov systems. More recently, Yin \textit{et al.}~\cite{yin2018higher} investigated discontinuity-induced bifurcations (DIBs) in impact oscillators near grazing using a higher-order ZDM.

This article proposes a transverse discontinuity map (TDM) tailored for Filippov systems, aiming to address the limitations associated with the first-order saltation matrix. Consider an $n$-dimensional Filippov system where the state vector $\mathbf{x} \in \mathbb{R}^n$ evolves according to either $\dot{\mathbf{x}} = \mathbf{F}_i(\mathbf{x})$ or $\dot{\mathbf{x}} = \mathbf{F}_j(\mathbf{x})$. The time required for an orbit to reach the discontinuity boundary $\Sigma_{i,j}$ during an impact at $\mathbf{x} = \mathbf{x}_i(t_i) \in \Sigma_{i,j}$ is given by  
\begin{equation}\label{eq 1}
    \delta_1 = -\frac{\nabla H(\mathbf{x}_i)^T \cdot \mathbf{y}_-}{\nabla H(\mathbf{x}_i)^T \cdot \mathbf{F}_i(\mathbf{x_i})},
\end{equation}
where $\mathbf{y}_-$ denotes the perturbed state during the impact. The governing vector fields in the subspaces $S_i$ and $S_j$ are represented by $\mathbf{F}_i(\mathbf{x})$ and $\mathbf{F}_j(\mathbf{x})$, respectively. The scalar function $H(\mathbf{x})$ defines the discontinuity surface as $\Sigma_{i,j} = \{\mathbf{x}_i \in \mathbb{R}^n : H(\mathbf{x}) = 0\}$.

The expression for flight time $\delta_1$ arises from a first-order Taylor series approximation of the discontinuity surface. However, this approximation presents a fundamental issue: since $\delta_1$ is a real-valued expression, it implies that all nearby orbits in the local neighborhood of the discontinuity are mapped to a common image given by $\mathbf{y}_+ = \mathbf{S}_1 \cdot \mathbf{y}_-$, where $\mathbf{S}_1$ denotes the first-order saltation matrix. Moreover, in the vicinity of low-velocity impacts near grazing, the denominator in Eq.~\eqref{eq 1}, $\nabla H(\mathbf{x}_i)^T \cdot \mathbf{F}_i(\mathbf{x}_i)$, approaches zero, leading to an overestimation of the mapped state $\mathbf{y}_+$. Consequently, the appropriate conditions under which the TDM, ZDM, or PDM should be employed in Filippov systems remain an open question.

This article demonstrates that certain perturbed orbits may fail to intersect the discontinuity boundary. This aspect cannot be accurately captured using the first-order transverse discontinuity map (TDM) or the associated saltation matrix. To overcome this limitation, a higher-order approximation of the TDM is introduced, leading to a quadratic equation in the flight time $\delta$. The emergence of a quadratic form in $\delta$ implies that perturbations reach the discontinuity boundary and are accurately mapped to $\mathbf{y} _+$ when the discriminant of the quadratic expression is positive.

Note, the positive root of the quadratic equation incorporates additional correction terms in the denominator, which prevent divergence of the mapped state during low-velocity impacts and avoid overestimation near grazing incidence. These crucial correction terms are absent in the first-order saltation matrix.

Such limitations have largely been overlooked in the literature, which primarily emphasizes discontinuity mappings such as the zero-time discontinuity mapping (ZDM) and the Poincar\'e discontinuity mapping (PDM) under grazing conditions~\cite{yin2018higher}. Furthermore, the higher-order correction to the TDM proposed in~\cite{bernardo2008piecewise} still relies on the first-order flight time $\delta_1$, which inherently retains the singularity in the denominator. This issue can be resolved by adopting a quadratic approximation in $\delta$ as presented in~\cite{chawla2023higher,chawla2025higher} for impact oscillators subjected to impacts with a rigid barrier.

One of the key challenges introduced by the higher-order transverse discontinuity map (TDM) is that the Jacobian matrix at the point of impact cannot be expressed in a closed-form analytical expression, unlike the first-order saltation matrix. This difficulty stems from the fact that the higher-order mapping of the post-impact state $\mathbf{y}_+$ involves nonlinear terms of the order $\mathcal{O}(\delta_+^2, \delta_+ \mathbf{y}, \mathbf{y}_- \cdot \mathbf{y}_-)$.

To overcome this limitation, the present study develops a numerical framework to evaluate the Jacobian matrix during impact, thereby enabling the computation of a higher-order saltation matrix. The validity of this proposed matrix is demonstrated by estimating the monodromy matrix of a piecewise-smooth (PWS) Filippov system and subsequently computing the corresponding Floquet multipliers and Lyapunov exponents.

This study also proposes an electronic analog of a soft-impacting system that provides experimental validation of discontinuity-induced bifurcations (DIBs) observed in a representative Filippov system. The designed circuit accurately reproduces the dynamical behavior of a bilinear oscillator across a wide range of stiffness ratios.

Previously, Seth and Banerjee \cite{seth2020electronic} introduced an equivalent electrical circuit composed of a series LCR circuit having a parallel capacitive branch controlled with a switching mechanism to achieve a similar objective. The LCR circuit serves as an analog of a mass-spring-damper system, where the parallel capacitive branch resembles a spring connected to a massless cushion in a mechanical impact system~\cite{banerjee2009invisible}. Their experimental results demonstrated the presence of narrow-band chaos and revealed the well-known finger-shaped chaotic attractor in the Poincar\'e plane under varying stiffness ratios $\beta$. However, their circuit design fails to accurately capture the bifurcation behavior when the stiffness ratio is significantly low. The underlying issue arises when $\beta$ is small, as the capacitance of the series LCR circuit matches that of the parallel capacitive branch. Consequently, the time required for the capacitor to discharge becomes substantial, preventing the parallel capacitor from discharging sufficiently within a short interval. This limitation prevents the circuit from operating effectively under very low $\beta$ values. To overcome this challenge, the proposed circuit design employs operational amplifiers to integrate two first-order ordinary differential equations (ODEs), effectively modeling a bilinear oscillator for low and high stiffness ratios. Additionally, this circuit is effective in observing coexisting multiple stable attractors, which is not possible in a series LCR circuit. In this work, we present experimentally obtained phase portraits, bifurcation diagrams, and finger-shaped Poincar\'e sections, all of which confirm the occurrence of DIBs, aligning well with numerical predictions derived from the higher-order TDM approach.

The computed results presented in this paper exhibit strong agreement with DIB diagrams obtained through both numerical simulations and experimental measurements, thereby confirming the effectiveness of the proposed approach.

The remainder of this paper is organized as follows. In Section~\ref{sec 2}, we introduce the higher-order TDM for Filippov systems. Section~\ref{sec 3} provides a comparative analysis of the higher-order TDM with the conventional first-order saltation matrix. In Section~\ref{sec 4}, we present a computational framework for numerically determining the higher-order saltation matrix, along with its associated Floquet multipliers and Lyapunov exponents. Section~\ref{sec 5} describes the design and implementation of an electronic analog circuit that serves as an experimental counterpart to the soft impact oscillator examined in Section~\ref{sec 3}, thereby enabling the experimental validation of discontinuity-induced bifurcations. Finally, Section~\ref{sec 6} summarizes the key findings of this study.

\section{Transverse discontinuity map} \label{sec 2}

Consider a vector $\mathbf{x} \in \mathbb{R}^n$ representing an arbitrary piecewise-smooth (PWS) Filippov system of dimension $n$. The components $x_1, x_2, \dots, x_n$ correspond to the measurable physical quantities of the system being modeled. The dynamics of the state vector $\mathbf{x}$ are governed by the differential equation
\begin{equation} \label{eq 2}
	\dot{\mathbf{x}} = 
	\begin{cases}
		\mathbf{F}_i(\mathbf{x}), & \text{if } \mathbf{x} \in S_i, \\
		\mathbf{F}_j(\mathbf{x}), & \text{if } \mathbf{x} \in S_j,
	\end{cases}
\end{equation}
where $S_i$ and $S_j$ partition the phase space into distinct subsets, each governed by different vector fields $\mathbf{F}_i(\mathbf{x})$ or $\mathbf{F}_j(\mathbf{x})$, respectively. For non-autonomous systems with explicit time dependence, \textit{i.e.}, $\mathbf{F}_i(\mathbf{x}, t)$, a new phase variable $s = t$ can be introduced such that $\dot{s} = 1$. By augmenting the state vector $\mathbf{x}$ with the additional variable $s$, the extended state vector $\mathbf{x}^\prime \in \mathbb{R}^{n+1}$ accounts for the time evolution explicitly. As a result, the subsequent analysis applies uniformly to both autonomous and non-autonomous systems.

To analyze the behavior of trajectories in the vicinity of a periodic orbit $\mathbf{x}_0$, consider a nearby (secondary) orbit given by $\tilde{\mathbf{x}} = \mathbf{x}_0 + \mathbf{y} \in S_i$, where $\mathbf{y}$ represents a small perturbation around $\mathbf{x}_0$. Substituting $\tilde{\mathbf{x}}$ into Eq.~\eqref{eq 2} and performing a Taylor series expansion of $\mathbf{F}(\mathbf{x})$ about $\mathbf{x}_0$, the dynamics of the perturbation $\mathbf{y}$ is governed by:

\begin{align} \label{eq 3}
    \dot{\mathbf{x}}_0 + \dot{\mathbf{y}} &= \mathbf{F}_i(\mathbf{x}_0) + \nabla \mathbf{F}_i(\mathbf{x}_0) \cdot \mathbf{y} + \mathcal{O}(\|\mathbf{y}\|^2), \\
    \dot{\mathbf{y}} &\approx \nabla \mathbf{F}_i(\mathbf{x}_0) \cdot \mathbf{y}. \nonumber
\end{align}
where $\dot{( )}$ denotes the derivative with respect to time $t$, $\nabla\mathbf{F}_i^T(\mathbf{x})$ is the Jacobian matrix of $\mathbf{F}_i(\mathbf{x})$, and $\dot{\mathbf{x}}_0 = \mathbf{F}_i(\mathbf{x}_0)$ follows from Eq.~\eqref{eq 2}. The evolution of $\mathbf{x}$ and $\tilde{\mathbf{x}}$ is piecewise-smooth, and they intersect the discontinuity boundary $\Sigma_{i,j}$, which separates regions $S_i$ and $S_j$, at different time instants.

\begin{figure}[tbh]
	\centering
	\includegraphics[scale = 0.7]{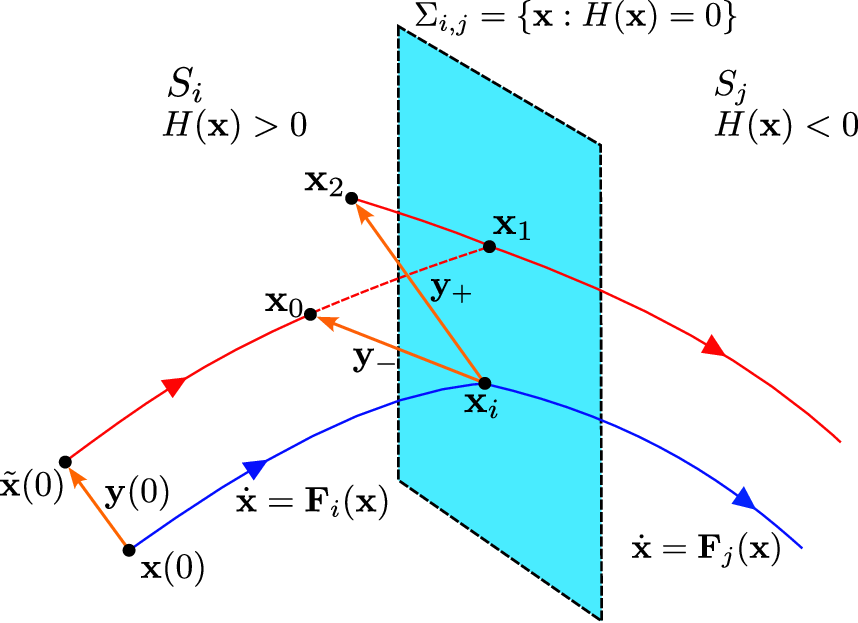}
	\caption{Schematic of two nearby trajectories $\mathbf{x}$ and $\tilde{\mathbf{x}} = \mathbf{x} + \mathbf{y}$ of a Fillipov system; shown as blue and red curves.} 
	\label{fig 1}
\end{figure}
To derive a closed-form expression for this time difference, consider the phase portrait of a two-dimensional system illustrated in Fig.~\ref{fig 1}, depicting the evolution of trajectories. Two nearby trajectories, $\mathbf{x}(0)$ and $\tilde{\mathbf{x}}(0) = \mathbf{x}(0) + \mathbf{y}(0)$, are initiated from region $S_i$. Initially, $\mathbf{x}(0)$ corresponds to a periodic solution of Eq.~\eqref{eq 2}, while $\tilde{\mathbf{x}}$ represents a perturbed orbit around $\mathbf{x}_0$, with the perturbation $\mathbf{y}(0)$.

As the trajectories evolve over time, $\mathbf{x}(0)$ reaches the point $\mathbf{x}(t_i) = \mathbf{x}_i$ on the discontinuity boundary $\Sigma_{i,j} := S_i \cap S_j \subset \mathbb{R}^{n-1}$ at time $t = t_i$. The boundary $\Sigma_{i,j}$ is defined by a scalar function $H(\mathbf{x}) \in \mathbb{R}$ such that $\Sigma_{i,j} = \{\mathbf{x} \in \mathbb{R}^n : H(\mathbf{x}) = 0\}$. Consequently, $S_i = \{\mathbf{x} \in \mathbb{R}^n : H(\mathbf{x}) > 0\}$ and $S_j = \{\mathbf{x} \in \mathbb{R}^n : H(\mathbf{x}) < 0\}$. At $t = t_i$, the governing vector field switches from $\mathbf{F}_i(\mathbf{x})$ to $\mathbf{F}_j(\mathbf{x})$.

However, at this same instant, the perturbed trajectory evolves to $\hat{\mathbf{x}}(t_i) = \mathbf{x}_0 \in S_i$, has not yet reached $\Sigma_{i,j}$. Therefore, integrating $\mathbf{x}_0$ forward using the Jacobian of $\mathbf{F}_j(\mathbf{x})$ is incorrect, as $\mathbf{x}_0 \in S_i$ will intersect $\Sigma_{i,j}$ later, at time $t = t_i + \delta$. Here, $\delta$ represents the flight time required for $\mathbf{x}_0$ to reach $\mathbf{x}_1 \in \Sigma_{i,j}$ (see Fig.~\ref{fig 1}).

To account for the discontinuities in the vector field near $\Sigma_{i,j}$, a Transverse Discontinuity Mapping (TDM) \cite{chawla2022stability,chawla2023higher,chawla2025higher} is employed. This mapping, which is fundamentally dependent on $\delta$, maps the state $\mathbf{x}_0$ to $\mathbf{x}_2$ such that the time taken by $\mathbf{x}_0 \in S_i$ to reach $\mathbf{x}_1 \in \Sigma_{i,j}$ equals the time taken by $\mathbf{x}_2 \in S_j$ to reach $\mathbf{x}_1$, now governed by the Jacobian of $\mathbf{F}_j(\mathbf{x})$. This ensures a zero-time mapping of the perturbed trajectories, preserving temporal consistency.

Once the state is mapped from $\mathbf{x}_0$ to $\mathbf{x}_2$ using the TDM, the corresponding perturbations are transformed from $\mathbf{y}_-$ to $\mathbf{y}_+$ using the Jacobian of the TDM. Under a linearized approximation, perturbations are typically mapped using a state transition matrix, known as the saltation matrix. A higher-order approximation for both the time difference $\delta$ and the TDM for the perturbed state $\hat{\mathbf{x}}$ and its perturbation $\mathbf{y}$ during an intersection with $\Sigma_{i,j}$ is derived in the following section.

A Taylor series expansion of the state vector $\mathbf{x}(t)$ and the vector field $\mathbf{F}(\mathbf{x})$, satisfying the dynamical equation $\dot{\mathbf{x}} = \mathbf{F}(\mathbf{x})$, about the time instant $t = t_0$ with $\mathbf{x}(t_0) = \mathbf{x}_0$, yields the following expressions:

\begin{align} \label{eq 4}
	\mathbf{x}(t) &= \mathbf{x}(t_0) + \Delta t \, \mathbf{F}(\mathbf{x}_0) + \dfrac{1}{2}\Delta t^2 \, \nabla \mathbf{F}(\mathbf{x}_0)^T \cdot \mathbf{F}(\mathbf{x}_0) + \mathcal{O}(\Delta t^3),
\end{align}

\begin{align} \label{eq 5}
	\mathbf{F}(\mathbf{x}) &= \mathbf{F}(\mathbf{x}_0) + \nabla \mathbf{F}(\mathbf{x}_0)^T \cdot \Delta \mathbf{x} + \dfrac{1}{2}
	\begin{bmatrix}
		\Delta \mathbf{x}^T \cdot \text{He}_1 \cdot \Delta \mathbf{x} \\
		\Delta \mathbf{x}^T \cdot \text{He}_2 \cdot \Delta \mathbf{x}
	\end{bmatrix} + \mathcal{O}(\|\Delta \mathbf{x}\|^3),
\end{align}

Here, the matrices $\text{He}_i$ denote the Hessians of each component $f_i(\mathbf{x})$ of the vector field $\mathbf{F}(\mathbf{x})$, that is, $\text{He}_i$ is defined as the Jacobian of the gradient $\nabla f_i(\mathbf{x})$, or equivalently, $\nabla(\nabla f_i(\mathbf{x}))^T$. Also, $\Delta \mathbf{x} = \mathbf{x} - \mathbf{x}_0$ represents the deviation from the reference state. 

Let $\delta$ be the time required for the state $\mathbf{x}_0$ to evolve to $\mathbf{x}_1$; see Fig.~\ref{fig 1}. Since the primary trajectory initiated from $\mathbf{x}(0)$ reaches $\mathbf{x}(t_i) = \mathbf{x}_i$ at time $t = t_i$, the perturbed trajectory $\tilde{\mathbf{x}}(0)$ evolves to $\tilde{\mathbf{x}}(t_i) = \mathbf{x}_0$. Therefore, after an additional time $\delta$, $\tilde{\mathbf{x}}(t_i + \delta) = \mathbf{x}_1 \in \Sigma_{i,j}$. 

The state $\mathbf{x}_1$ can be approximated by performing a Taylor series expansion about $\mathbf{x}_0$ at time $t = \delta$, yielding:
\begin{equation} \label{eq 6}
	\mathbf{x}_1 = \mathbf{x}_0 + \delta \, \mathbf{F}_i(\mathbf{x}_0) + \frac{\delta^2}{2} \nabla \mathbf{F}_i^T(\mathbf{x}_0) \cdot \mathbf{F}_i(\mathbf{x}_0) + \mathcal{O}(\delta^3).
\end{equation}

Using the expansion from Eq.~\eqref{eq 5}, the vector field $\mathbf{F}_i(\mathbf{x}_0)$ can be approximated by performing a Taylor series expansion of $\mathbf{F}_i(\mathbf{x})$ about the primary impact state $\mathbf{x}_i$, and evaluating it at $\mathbf{x}_0 = \mathbf{x}_i + \mathbf{y}_-$ (see Fig.~\ref{fig 1}). Here, $\mathbf{y}_-$ denotes the perturbation of the trajectory $\tilde{\mathbf{x}}$ relative to the nominal state $\mathbf{x}_i$ at the moment of impact with the discontinuity boundary. 

Accordingly, the Taylor-expanded form of $\mathbf{F}_i(\mathbf{x}_0)$ becomes:
\begin{equation} \label{eq 7}
	\mathbf{F}_i(\mathbf{x}_0) = \mathbf{F}_i(\mathbf{x}_i) + \nabla \mathbf{F}_i^T(\mathbf{x}_i) \cdot \mathbf{y}_- + \frac{1}{2}\begin{pmatrix}
		\mathbf{y}_-^T \cdot \text{He}(f^i_1) \cdot \mathbf{y}_-\\
		\mathbf{y}_-^T \cdot \text{He}(f^i_2) \cdot \mathbf{y}_-
	\end{pmatrix} + \mathcal{O}(||\mathbf{y}_-||^3),
\end{equation}

Here, $\text{He}(f^i_1)$ and $\text{He}(f^i_2)$ denote the Hessian matrices corresponding to the scalar components $f^i_1(\mathbf{x})$ and $f^i_2(\mathbf{x})$ of the vector field $\mathbf{F}_i(\mathbf{x})$. Substituting the expression for $\mathbf{F}_i(\mathbf{x}_0)$ from Eq.~\eqref{eq 7} into the Taylor expansion in Eq.~\eqref{eq 6}, the state $\mathbf{x}_1$ can be approximated as:

\begin{align} \label{eq 8}
	\mathbf{x}_1 = \mathbf{x}_0 &+ \delta\left(\mathbf{F}_i(\mathbf{x}_i) + \nabla \mathbf{F}_i^T(\mathbf{x}_i) \cdot \mathbf{y}_-\right) + \frac{1}{2}\delta^2 \nabla \mathbf{F}_i^T(\mathbf{x}_i) \cdot \mathbf{F}_i(\mathbf{x}_i) \\
	&\quad + \mathcal{O}(\delta^3, ||\mathbf{y}_-||^3, \delta^2 ||\mathbf{y}_-||, ||\mathbf{y}_-||^2 \delta). \nonumber
\end{align}

The flight time $\delta$ required for the state $\mathbf{x}_0$ to reach $\mathbf{x}_1$ is determined by solving the condition $H(\mathbf{x}_1) = 0$. A Taylor expansion of the scalar function $H(\mathbf{x})$ about the primary impact state $\mathbf{x}_i$, evaluated at $\mathbf{x}_1$, yields:

\begin{equation} \label{eq 9}
	H(\mathbf{x}_1) = H(\mathbf{x}_i) + \nabla H(\mathbf{x}_i)^T \cdot (\mathbf{x}_1 - \mathbf{x}_i) + \frac{1}{2}(\mathbf{x}_1 - \mathbf{x}_i)^T \cdot \text{He}(H(\mathbf{x}_i)) \cdot (\mathbf{x}_1 - \mathbf{x}_i) + \mathcal{O}\left(\| \mathbf{x}_1 - \mathbf{x}_i \|^3 \right).
\end{equation}

Since both $\mathbf{x}_1$ and $\mathbf{x}_i$ lie on the discontinuity boundary $\Sigma_{i,j}$, it follows that $H(\mathbf{x}_1) = H(\mathbf{x}_i) = 0$, due to the definition $\Sigma_{i,j} = \{\mathbf{x} \in \mathbb{R}^n : H(\mathbf{x}) = 0\}$. Substituting Eq.~\eqref{eq 8} into Eq.~\eqref{eq 9}, along with the relation $\mathbf{x}_0 = \mathbf{x}_i + \mathbf{y}_-$, an analytical expression for the flight time $\delta$ can be obtained. This results in a quadratic equation in $\delta$, as expressed in Eq.~\eqref{eq 10}.

\begin{align}\label{eq 10}
	\delta^2\Big(&\nabla H(\mathbf{x}_i)^T \cdot \nabla \mathbf{F}_i(\mathbf{x}_i)^T \cdot \mathbf{F}_i(\mathbf{x}_i) + \mathbf{F}_i(\mathbf{x}_i)^T \cdot \nabla (\nabla H(\mathbf{x}_i))^T \cdot \mathbf{F}_i(\mathbf{x}_i)\Big) \nonumber \\
	&+ \delta \Big( 2 \nabla H(\mathbf{x}_i)^T \cdot \mathbf{F}_i(\mathbf{x}_i) + 2 \nabla H(\mathbf{x}_i)^T \cdot \nabla \mathbf{F}_i(\mathbf{x}_i)^T \cdot \mathbf{y}_- \nonumber \\
	&\quad+ \mathbf{y}_-^T \cdot \nabla (\nabla H(\mathbf{x}_i))^T \cdot \mathbf{F}_i(\mathbf{x}_i) + \mathbf{F}_i(\mathbf{x}_i)^T \cdot \nabla (\nabla H(\mathbf{x}_i))^T \cdot \mathbf{y}_- \Big) \nonumber \\
	&\qquad+ \mathbf{y}_-^T \cdot \nabla (\nabla H(\mathbf{x}_i))^T \cdot \mathbf{y}_- + 2 \nabla H(\mathbf{x}_i)^T \cdot \mathbf{y}_- \nonumber \\
	&\quad \qquad + \mathcal{O}(3) = G(\delta, \mathbf{x}_i, \mathbf{y}_-) = 0,
\end{align}
where the quadratic expression in $\delta$ is denoted by $G(\delta, \mathbf{x}_i, \mathbf{y}_-)$. By retaining only the linear terms in $\mathbf{y}_-$, a first-order approximation of the flight time, denoted as $\delta_1$, can be obtained as follows:

\begin{equation} \label{eq 11}
	\delta_1 = -\dfrac{\nabla H(\mathbf{x}_i)^T \cdot \mathbf{y}_-}{\nabla H(\mathbf{x}_i)^T \cdot \mathbf{F}_i(\mathbf{x}_i)}.
\end{equation}

Since $G(\delta, \mathbf{x}_i, \mathbf{y}_-)$ is a quadratic expression in $\delta$, Eq.~\eqref{eq 10} can be reformulated in the standard quadratic form as
\begin{equation} \label{eq 12}
	G(\delta, \mathbf{x}_i, \mathbf{y}_-) = A \delta^2 + B \delta + C,
\end{equation}
where the coefficients $A$, $B$, and $C$ are scalar quantities defined as follows:

\begin{align} \label{eq 13}
	A &= \mathbf{\nabla}H(\mathbf{x}_i)^T\cdot\mathbf{\nabla}\mathbf{F_i}(\mathbf{x}_i)^T\cdot\mathbf{F_i}(\mathbf{x}_i) + \mathbf{F_i}(\mathbf{x}_i)^T\cdot\mathbf{\nabla}(\mathbf{\nabla}H(\mathbf{x}_i))^T\cdot\mathbf{F_i}(\mathbf{x}_i),\\
	B &= 2\mathbf{\nabla} H(\mathbf{x}_i)^T\cdot\mathbf{F_i}(\mathbf{x}_i) + 2\mathbf{\nabla}H(\mathbf{x}_i)^T\cdot\mathbf{\nabla}\mathbf{F_i}(\mathbf{x}_i)^T\cdot\mathbf{y}_- \nonumber \\
	&\quad + \mathbf{y}_-^T\cdot\mathbf{\nabla}(\mathbf{\nabla}H(\mathbf{x}_i))^T\cdot\mathbf{F_i}(\mathbf{x}_i) + \mathbf{F_i}(\mathbf{x}_i)^T\cdot\mathbf{\nabla}(\mathbf{\nabla}H(\mathbf{x}_i))^T\cdot\mathbf{y}_-, \nonumber \\
	C &= \mathbf{y}_-^T\cdot\mathbf{\nabla}(\mathbf{\nabla}H(\mathbf{x}_i))^T\cdot\mathbf{y}_- + 2 \mathbf{\nabla}H(\mathbf{x}_i)^T\cdot \mathbf{y}_-. \nonumber
\end{align}

The root of the equation $G(\delta, \mathbf{x}_i, \mathbf{y}_-) = 0$ or equivalently $H(\mathbf{x}_1) = 0$ implies that nearby trajectories undergo impacts with the set $\Sigma_{i,j}$ at the time $t = t_i + \delta$. A comparison of Eqs.~\eqref{eq 10} and \eqref{eq 11} raises two important questions. First, is it possible for imaginary roots of Eq. \eqref{eq 10} to exist, implying that no impacts will occur, a scenario that the first-order approximation of $\delta_1$ (i.e., Eq. \eqref{eq 11}) does not capture? Second, should the perturbation $\mathbf{y}_-$ lie within a critical range such that impacts will indeed occur? 

The higher-order estimate for $\delta$ suggests that perturbations larger than a critical threshold will not result in impacts. This occurs when the discriminant of Eq.~\eqref{eq 10} is negative, i.e., when $B^2 - 4AC \leq 0$, resulting in $\delta \in \mathbb{C}$. The correct root of $G(\delta, \mathbf{x}_i, \mathbf{y}_-) = 0$ that determines the flight time of impact in the local neighborhood of $\mathbf{x}_i$ is given by the positive root:

\begin{align} \label{eq 14}
	\delta_+ = -\frac{B}{2A} \left( 1 - \sqrt{1 - \frac{4AC}{B^2}} \right).
\end{align}

This ensures that $\delta_+ \to 0$ as $\mathbf{y}_- \to 0$. The critical range in $\mathbf{y}_-$, beyond which no impacts occur and perturbations will miss the discontinuity barrier $\Sigma_{i,j}$, is given by the condition $B^2 \geq 4AC$. Consequently, when $\delta_+$ is imaginary, no real roots exist that satisfy $H(\mathbf{x}_1) = 0$, meaning that perturbations do not reach and impact the discontinuity barrier $\Sigma_{i,j}$. In contrast, the first-order approximation $\delta = \delta_1$ yields a real value, predicting that impacts occur for all perturbations. The higher-order approach, however, provides an upper limit for when the first-order saltation matrix breaks down for low-velocity impacts near grazing. This phenomenon has been demonstrated in Section \ref{sec 3}.

Next, a Taylor expansion of $\mathbf{x}$ about the point $\mathbf{x}(0) = \mathbf{x}_1$, evaluated backward in time at $t = -\delta_+$, gives the mapped state $\mathbf{x}_2$:
\begin{align} \label{eq 15}
	\mathbf{x}_2 = \mathbf{x}_1 - \delta_+ \mathbf{F}_j(\mathbf{x}_1) + \frac{\delta_+^2}{2}\nabla \mathbf{F}_j^T(\mathbf{x}_1) \cdot \mathbf{F}_j(\mathbf{x}_1) + \mathcal{O}(\delta_+^3).
\end{align}
The vector field $\mathbf{F}_j(\mathbf{x}_1)$ is approximated by expanding $\mathbf{F}_j(\mathbf{x})$ around $\mathbf{x}_i$ and evaluating it at $\mathbf{x}_1$. From Eq.~\eqref{eq 8} and the relation $\mathbf{x}_0 = \mathbf{x}_i + \mathbf{y}_-$, $\mathbf{F}_j(\mathbf{x}_1)$ can be expressed up to $\mathcal{O}(1)$ as:
\begin{equation} \label{eq 16}
	\mathbf{F}_j(\mathbf{x}_1) = \mathbf{F}_j(\mathbf{x}_i) + \nabla \mathbf{F}_j^T \cdot (\mathbf{y}_- + \delta_+ \mathbf{F}_i(\mathbf{x}_i)) + \mathcal{O}(||\mathbf{y}_-||^2, \delta_+ ||\mathbf{y}_-||, \delta_+^2).
\end{equation}

Finally, by substituting Eqs. \eqref{eq 8} and \eqref{eq 16} into Eq. \eqref{eq 15} and using the positive root $\delta_+$, the time-dependent map (TDM) from $\mathbf{x}_0$ to $\mathbf{x}_2$ can be expressed in terms of $\mathbf{y}_-$ and $\mathbf{y}_+$ as follows:
\begin{align} \label{eq 17}
	\mathbf{y}_+ = \mathbf{y}_- &+ \delta \Big( \mathbf{F}_i(\mathbf{x}_i) - \mathbf{F}_j(\mathbf{x}_i) \Big) + \delta \Big( \nabla \mathbf{F}_i^T(\mathbf{x}_i) - \nabla \mathbf{F}_j^T(\mathbf{x}_i) \Big) \cdot \mathbf{y}_- \nonumber \\
	&\quad + \frac{\delta^2}{2} \Big( \nabla \mathbf{F}_i^T(\mathbf{x}_i) \cdot \mathbf{F}_i(\mathbf{x}_i) - 2 \nabla \mathbf{F}_j^T(\mathbf{x}_i) \cdot \mathbf{F}_i(\mathbf{x}_i) + \nabla \mathbf{F}_j^T(\mathbf{x}_i) \cdot \mathbf{F}_j(\mathbf{x}_i) \Big) \nonumber \\
	&\qquad + \mathcal{O}(\delta^3, \delta^2 ||\mathbf{y}_-||, \delta ||\mathbf{y}_-||^2).
\end{align}

where $\mathbf{x}_2 = \mathbf{x}_i + \mathbf{y}_+$; see Fig.~\ref{fig 1}. Note that by keeping only the linear order terms in Eq.~\eqref{eq 17} and using the $\mathcal{O}(1)$ approximation of $\delta$ (see Eq.~\eqref{eq 11}), the mapping from $\mathbf{y}_-$ to $\mathbf{y}_+$ can be expressed in matrix form as,
\begin{equation} \label{eq 18}
	\mathbf{y}_+ = \mathbf{S}_1 \cdot \mathbf{y}_-
\end{equation}
where $\mathbf{S}_1$ is the saltation matrix \cite{leine2000bifurcations} for Filippov systems, and is given by,

\begin{equation} \label{eq 19}
	\mathbf{S}_1 = \mathbb{I}_{n \times n} + \frac{\big( \mathbf{F}_j(\mathbf{x}_i) - \mathbf{F}_i(\mathbf{x}_i) \big)}{\nabla H(\mathbf{x}_i)^T \cdot \mathbf{F}_i(\mathbf{x}_i)} \otimes \nabla H(\mathbf{x}_i).
\end{equation}
The TDM, as defined by Eqs.~\eqref{eq 10} and \eqref{eq 17}, includes higher-order corrections to the first-order saltation matrix, which is given in Eqs.~\eqref{eq 11} and \eqref{eq 19}. The proposed higher-order TDM is capable of accurately predicting impact occurrences within the local neighborhood of the discontinuity barrier, an ability that the linearized saltation matrix sometimes fails to capture. This is numerically illustrated in Sec .~\ref {sec 3} for a pre-stressed impact oscillator performing soft impacts with an elastic barrier.

\section{Numerical results} \label{sec 3}

\begin{figure}[tbh]
    \centering
    \includegraphics[scale=1.3]{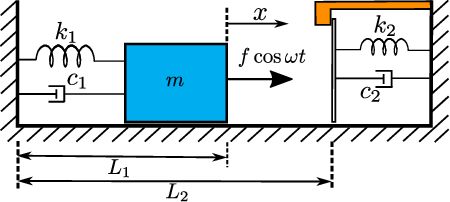}
    \caption{Schematic representation of a bilinear oscillator interacting with a pre-stressed deformable barrier.}
    \label{fig 2}
\end{figure}
This section demonstrates the limitations of the first-order saltation matrix (\textit{i.e.}, Eqs.~\eqref{eq 11} and \eqref{eq 19}) for low-velocity impacts. Furthermore, the inclusion of additional correction terms in the higher-order TDM enhances the accuracy of the mapped state compared to the first-order saltation matrix. To illustrate this, a periodically excited bilinear oscillator undergoing soft impacts with a deformable barrier is shown in Fig.~\ref{fig 2}.

A mass $m = 1$~kg is elastically attached to a fixed wall, as illustrated in Fig.~\ref{fig 2}. The mass is subjected to a sinusoidal driving force with an amplitude of $f$ and a forcing frequency of $\omega = 0.8$~Hz, causing it to oscillate on a frictionless surface. The mass $m$ comes into contact with a pre-stressed barrier when the relative displacement exceeds $L_2 - L_1 = 1.5$~m. At this instant, an additional restoring force is exerted on the mass due to the resistance of the barrier, which persists until the contact ceases. The oscillatory dynamics of $m$ can be described by a bilinear spring model with spring constants $k_1 = k_2 = 1.0$~N/m and damping coefficients $c_1 = c_2 = 0.1$~Ns/m.

The governing equations of motion describe a piecewise-smooth (PWS) dynamical system (see Eq.~\eqref{eq 2}) and are given by  
\begin{align} \label{eq 20}
    m \ddot{x} = 
    \begin{cases}
        f \cos{\omega t} - k_1 x  - c_1 \dot{x}, \quad & x < L_2 - L_1, \\
        f \cos{\omega t} - (k_1  + k_2)x - (c_1 + c_2) \dot{x}, \quad & x \geq L_2 - L_1.
    \end{cases}
\end{align}

The orbital divergence of limit cycles is analyzed by numerically integrating Eqs.~\eqref{eq 20} in \textit{Mathematica} with an absolute tolerance of up to $15^{\text{th}}$ decimal precision. For the selected system, the discontinuity boundary $\Sigma_{i,j}$ is located at $x = 1.5$~m, as determined by the discontinuity function $H(\mathbf{x}) = x - (L_2 - L_1)$. To eliminate transient effects, the system is evolved for 300 oscillation cycles, each of period $2\pi/\omega$, before recording the results.

To characterize the topology of the observed limit cycles, the system state vector \( \mathbf{x}(t) \) is sampled stroboscopically at discrete-time intervals corresponding to integer multiples of the forcing period, i.e., when \( t \bmod T = 0 \), with \( T = 2\pi/\omega \). A periodic orbit is classified as a period-1 (P1) limit cycle if the trajectory returns to its initial state after one forcing period, \( T \). Similarly, if the state recurs after two forcing periods, \( 2T \), the orbit is identified as a period-2 (P2) limit cycle.

To investigate the local stability of a period-1 (P1) orbit observed at an excitation amplitude of \( f = 0.57~\text{N} \), the evolution of two trajectories in its vicinity is considered. Let \( \mathbf{x}(t) \) denote the reference trajectory on the P1 orbit, and \( \tilde{\mathbf{x}}(t) = \mathbf{x}(t) + \mathbf{y}(t) \) represent a perturbed trajectory, where \( \mathbf{y}(t) \) is the initial perturbation vector. The trajectories are initialized at time \( t = 330813~\text{s} \) with the following initial conditions:
\[
\mathbf{x}(330813) = \{-1.275, -0.536\}, \quad \mathbf{y}(330813) = \frac{0.1}{\sqrt{2}} \{1, 1\}.
\]

At time $t_i = 330818$~s, the trajectory $\mathbf{x} = \mathbf{x}_i$ undergoes an impact with the discontinuity boundary $\Sigma_{i,j}$, while the perturbation norm is given by $||\mathbf{y}_-|| = 0.0842268$. At this instant, the higher-order flight time $\delta$ for the perturbed trajectory $\tilde{\mathbf{x}}$ to reach $\Sigma_{i,j}$ can be determined by solving the equation $G(\delta, \mathbf{x}_i, \mathbf{y}_-) = 0$, which is obtained from Eq.~\eqref{eq 10},

At time \( t_i = 330818~\text{s} \), the trajectory satisfies \( \mathbf{x}(t_i) = \mathbf{x}_i \), and undergoes an impact with the discontinuity boundary \( \Sigma_{i,j} \). At this instant, the perturbation norm associated with the nearby trajectory is given by \( \|\mathbf{y}_-\| = 0.0842268 \). The corresponding higher-order flight time \( \delta \), representing the time required for the perturbed trajectory \( \tilde{\mathbf{x}} = \mathbf{x} + \mathbf{y} \) to reach the boundary \( \Sigma_{i,j} \), can be computed by solving the nonlinear equation
\[
G(\delta, \mathbf{x}_i, \mathbf{y}_-) = 0,
\]
as derived from Eq.~\eqref{eq 10} below:
\begin{equation} \label{eq 21}
    \delta^2 \left(f \cos{\omega t} - k_1 \sigma - c_1 v \right) + 2\delta m (v + y_2) + 2m y_1 = 0,
\end{equation}
where the state vector at impact is given by \( \mathbf{x}_i = \{\sigma, v\} \), and the components of the perturbation vector \( \mathbf{y}_- \) are denoted as \( y_1 \) and \( y_2 \). These represent the respective displacements in position and velocity directions. 

\begin{figure}[tbh]
\centering
\includegraphics[scale = 0.55]{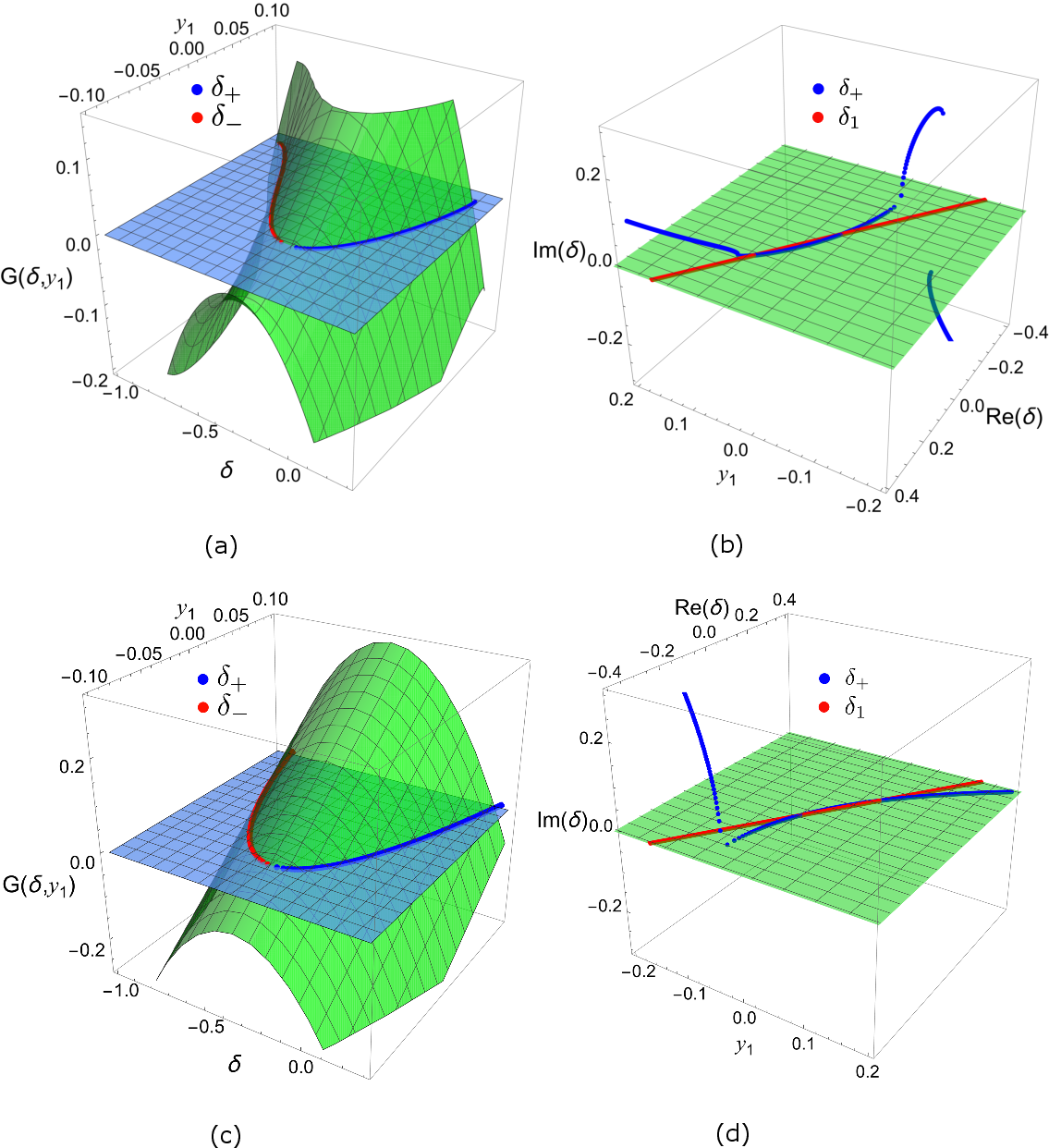}
\caption{Surface plots of \( G(\delta, y_1) \) as a function of \( \delta \) and \( y_1 \), for a forcing amplitude \( f = 0.57~\mathrm{N} \), evaluated at \( \mathbf{x}_i = \{1.5, -0.349336\} \) and perturbation vector \( \mathbf{y}_- = \{-0.0810259, 0.02299\} \), with norm \( \|\mathbf{y}_-\| = 0.0842268 \). In panel (a), the perturbation norm is held fixed at \( \|\mathbf{y}_-\| = 0.0842268 \), whereas in panel (c), the velocity component is fixed at \( y_2 = 0.02299 \). The blue plane corresponds to the zero-level set defined by \( G(\delta, y_1) = 0 \). The intersections of the surface with this plane, shown as red and blue points, represent the negative and positive real roots, respectively, that satisfy the impact condition. Panels (b) and (d) display the real and imaginary parts of the corresponding flight time \( \delta \) as a function of \( y_1 \), under the same constraints as in panels (a) and (c), respectively. Red and blue markers indicate the first-order (\textit{i.e.}, \( \delta_1 \)) and higher-order (\textit{i.e.}, \( \delta_+ \)) estimates of the flight time.}
\label{fig 3}
\end{figure}

Fig.~\ref{fig 3}(a) displays a surface plot of the function \( G(\delta, y_1) \), evaluated over the domain \( -1.0 \leq \delta \leq 0.3 \) and \( -0.1 \leq y_1 \leq 0.1 \). The values of \( y_1 \) and \( y_2 \) are chosen such that the norm of the perturbation vector remains constant at \( \|\mathbf{y}_-\| = 0.0842268 \) at the time of impact.

The blue and red points in Fig.~\ref{fig 3}(a) correspond to the positive and negative roots satisfying $G(\delta_+, y_1) = 0$ and $G(\delta_-, y_1) = 0$, respectively, as given by Eq.~\eqref{eq 22}. Consequently, the solutions incorporating higher-order corrections lie on the blue plane and accurately predict the flight time required for the perturbed trajectory $\tilde{\mathbf{x}}$ to reach the discontinuity boundary $\Sigma_{i,j}$.

Fig.~\ref{fig 3}(b) presents a comparison between the first-order estimate of the flight time, denoted by \( \delta_1 \) (as defined in Eq.~\eqref{eq 11}), and the real and imaginary components of the higher-order correction \( \delta_+ \), obtained by solving the nonlinear equation \( G(\delta_+, y_1) = 0 \). The solution \( \delta_+ \) corresponds to the positive root of the quadratic form derived from Eq.~\eqref{eq 20}. For this system, the first-order and higher-order expressions for the flight time are given by:

\begin{align} \label{eq 22}
    \delta_1 &= -\frac{y_1}{v}, \\
    \delta_+ &= \frac{v + y_2}{\alpha_1} \left( -1 + \sqrt{1 - \frac{2 \alpha_1 y_1}{(v + y_2)^2}} \right), \nonumber \\
    \delta_- &= \frac{v + y_2}{\alpha_1} \left( -1 - \sqrt{1 - \frac{2 \alpha_1 y_1}{(v + y_2)^2}} \right), \nonumber
\end{align}

where \( \delta_+ \) and \( \delta_- \) denote the two possible roots of the quadratic equation, and the coefficient \( \alpha_1 \) is defined as
\begin{equation}
    \alpha_1 = \frac{f}{m} \cos(\omega t) - \frac{k_1}{m} \sigma - \frac{c_1}{m} v.
\end{equation}

It is important to note that the trajectory originating from the region $S_i$ undergoes an impact at the discontinuity boundary $\Sigma_{i,j}$ before transitioning into the region $S_j$. To evaluate $\delta_+$ when orbits evolve from $S_j$ and subsequently cross $\Sigma_{i,j}$ from $S_j$ to $S_i$, the parameter $\alpha_1$ in Eq.~\eqref{eq 22} must be replaced with $\alpha_2$, where
\begin{equation}
    \alpha_2 = \frac{f}{m} \cos{\omega t} - \frac{(k_1 + k_2)}{m} \sigma - \frac{(c_1 + c_2)}{m} v.
\end{equation}

The higher-order results presented in Fig.~\ref{fig 3}(a) reveal that not all values of $y_1$ yield a real root, indicating that impacts do not occur for certain trajectories within the local neighborhood of $\mathbf{x}_i \in \Sigma_{i,j}$. The valid impact conditions are represented by the curve corresponding to the intersection of $G(\delta, y_1)$ with the constraint plane $G(\delta_+, y_1) = 0$. The locus of these impact points follows from Eq.~\eqref{eq 22}, with only the positive root being physically admissible, as it satisfies the condition $\delta_+ \rightarrow 0$ when $\mathbf{y}_- \rightarrow 0$.

A detailed investigation of $\delta_+$ in Fig.~\ref{fig 3}(b) further confirms that impacts occur only for specific values of $y_1$. This observation arises from the fact that only those solutions lying on the green plane, where the imaginary component of $\delta_+$ vanishes, hold physical significance, implying that impacts are realized in such instances. This result fundamentally contrasts with the predictions obtained from first-order approximations, where $\delta_1$ (Eq.~\eqref{eq 22}) remains a real-valued fraction, incorrectly suggesting that an impact occurs for all values of $y_1$.  

Therefore, the occurrence of impacts in locally perturbed trajectories is governed not only by the magnitude of the perturbation \( \|\mathbf{y}_-\| \) at the instant of impact at \( \mathbf{x}_i \), but also by the specific directional components of the perturbation vector \( \mathbf{y}_- \). This observation underscores the importance of incorporating higher-order corrections in order to faithfully capture the intricate dynamics of systems with discontinuous transitions.

Figs.~\ref{fig 3}(c) and (d) present a complementary analysis of the function \( G(\delta, y_1) \) and the associated flight time correction \( \delta_+ \), following the framework established in Figs.~\ref{fig 3}(a) and (b). In this case, the second component of the perturbation vector is held fixed at \( y_2 = 0.02299 \), as determined numerically from \( \mathbf{y}_- \), while the first component \( y_1 \) is varied over the range \( -0.1 \leq y_1 \leq 0.1 \). Consequently, the norm \( \|\mathbf{y}_-\| \) varies as \( y_1 \) changes independently. Fig.~\ref{fig 3}(c) illustrates a surface plot of \( G(\delta, y_1) \) as a function of \( \delta \) and \( y_1 \), spanning the domain \( -1.0 \leq \delta \leq 0.3 \) and \( -0.1 \leq y_1 \leq 0.1 \). The blue and red markers represent the positive and negative roots \( \delta_+ \) and \( \delta_- \), respectively, as defined in Eq.~\eqref{eq 22}.

The higher-order analysis reveals that impacts are not guaranteed for all perturbations. Specifically, trajectories in the vicinity of \( \mathbf{x}_i \) having component \( y_1 \) will only result in an impact with the boundary \( \Sigma_{i,j} \) if the condition \( G(\delta, y_1) = 0 \) is satisfied. These solutions lie at the intersection of the surface \( G(\delta, y_1) \) with the zero-level plane \( G(\delta_+, y_1) = 0 \), visualized as the blue plane in Fig.~\ref{fig 3}(c). The locus of these intersection points corresponds to the positive root \( \delta_+ \) from Eq.~\eqref{eq 22}.

Fig.~\ref{fig 3}(d) plots the real and imaginary components of the first-order (red) and higher-order (blue) flight times as functions of \( y_1 \). The higher-order analysis suggests that only those perturbations for which the imaginary part of \( \delta_+ \) vanishes—corresponding to intersections with the green plane—lead to actual impacts occurring at time \( t_i + \delta_+ \). In contrast, the first-order approximation incorrectly predicts that impacts occur for all values of \( y_1 \). This discrepancy emphasizes that not all trajectories within the local neighborhood of an impacting orbit will necessarily intersect the discontinuity boundary \( \Sigma_{i,j} \). The occurrence of impact is intricately linked not only to the perturbation magnitude but also to its directional components at the moment of impact.


\begin{figure}[tbh]
	\centering
	\includegraphics[scale = 0.55]{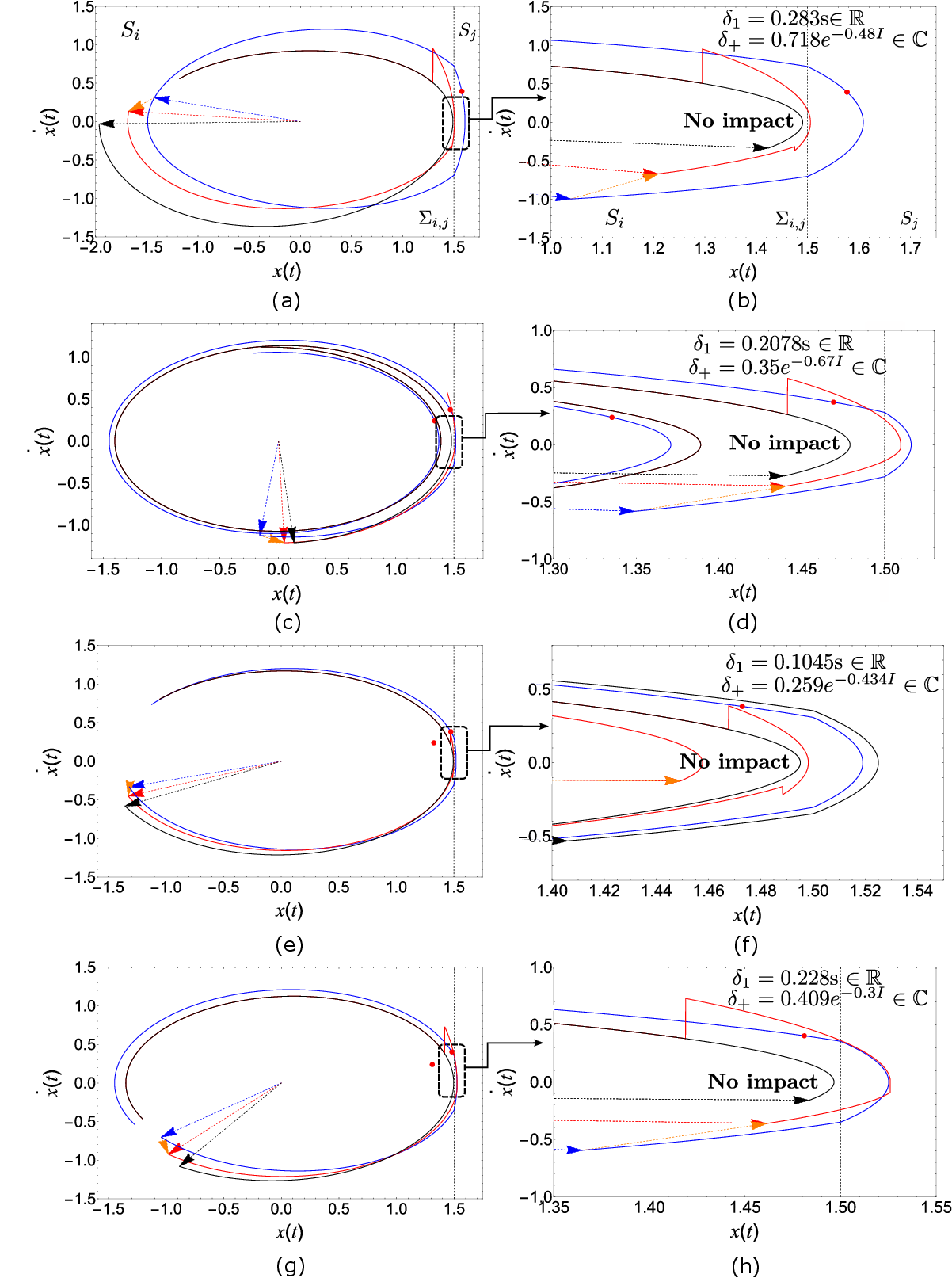}
	\caption{Phase portraits of the periodic orbit \( \mathbf{x} \) (blue) and perturbed orbit \( \mathbf{x} + \mathbf{y} \) (red) for (a) \( f = 0.7825 \, \mathrm{N} \), and (c), (e), (g) for \( f = 0.564 \), \( 0.566 \), and \( 0.57 \, \mathrm{N} \), respectively. (b), (d), (f), and (h) provide magnified views near the discontinuity boundary \( \Sigma_{i,j} \), highlighting the predicted impacts. First-order estimates (\( \delta_1 \in \mathbb{R} \)) indicate impacts, whereas higher-order corrections (\( \delta_+ \in \mathbb{C} \)) suggest no impact. True perturbed trajectories (black dashed) validate the higher-order predictions.}
	\label{fig 4}
\end{figure}
To validate the accuracy of the higher-order flight time predictions, we now compare the first-order analytical estimates with the actual numerical solutions of nearby trajectories. Fig.~\ref{fig 4} presents phase portraits illustrating the evolution of two neighboring trajectories for four different values of the forcing amplitude \( f \). In Fig.~\ref{fig 4}(a), the blue curve represents a period-1 (P1) limit cycle denoted by \( \mathbf{x} \) corresponding to \( f = 0.7825~\mathrm{N} \), while the red curve illustrates the perturbed trajectory \( \tilde{\mathbf{x}} = \mathbf{x} + \mathbf{y} \), initialized with the perturbation vector \( \mathbf{y} = \frac{0.35}{\sqrt{2}}\{1, 1\} \). The black curve indicates the true numerical solution obtained by integrating the full system dynamics from the initial condition \( \mathbf{x} + \mathbf{y} \).
Fig.~\ref{fig 4}(b) provides a magnified view of the phase portraits presented in Fig.~\ref{fig 4}(a), highlighting the behavior of both trajectories in the vicinity of the impact boundary \( \Sigma_{i,j} \). Based on the first-order approximation, the perturbed trajectory (shown in red) is expected to undergo an impact with \( \Sigma_{i,j} \) after a flight time of \( 0.283~\mathrm{s} \), resulting in a jump predicted by the first-order saltation matrix, synchronized with the impact of the primary orbit (shown in blue). This discontinuous transition is observed as the first-order trajectory deviation map (TDM) in Figs.~\ref{fig 4}(a) and (b).

However, the higher-order correction yields a complex flight time, \( \delta_+ = 0.718 e^{-0.48i} \in \mathbb{C} \), indicating the absence of a real root satisfying the impact condition \( H(\mathbf{x}_i + \mathbf{y}_-) = 0 \) at \( t_i + \delta_+ \). This implies that no physical impact should occur. This prediction is corroborated by the true numerical trajectory (depicted in black), which bypasses the discontinuity boundary \( \Sigma_{i,j} \) entirely.

A similar analysis is carried out in Figs.~\ref{fig 4}(c), (e), and (g), which show phase portraits of perturbed trajectories for excitation amplitudes \( f = 0.564 \), \( 0.566 \), and \( 0.57~\mathrm{N} \), respectively. The corresponding zoomed-in views near the impact location \( \mathbf{x}_i \) are shown in Figs.~\ref{fig 4}(d), (f), and (h). In each case, the first-order approximation predicts finite flight times (denoted \( \delta_1 \), see Eq.~\eqref{eq 22}) and suggests that impacts occur. However, the higher-order correction again yields complex values of \( \delta_+ \), implying the non-occurrence of impacts. This is confirmed by the true numerical solutions (black trajectories), which consistently avoid intersection with \( \Sigma_{i,j} \).

It is worth noting that the predictions of Fig.~\ref{fig 3} anticipated the absence of impacts for \( f = 0.57~\mathrm{N} \), which is validated in Figs.~\ref{fig 4}(g) and (h) through direct numerical simulation.

A comparison between the first-order and higher-order expressions for flight time, \( \delta_1 \) and \( \delta_+ \) in Eqs.~\eqref{eq 22}, reveals that the higher-order correction effectively regulates the flight time estimation near grazing impacts. Specifically, the presence of the correction terms ensures that the denominator in \( \delta_+ \) does not become arbitrarily small in the limit of low-impact velocities, thereby preventing divergence of the mapping near grazing conditions. This avoids overestimating the mapped state on the discontinuity boundary.

Importantly, the higher-order corrections incorporate key system parameters such as stiffness \( k_1 \), damping \( c_1 \), mass \( m \), and the external forcing parameters \( f \) and \( \omega \)—factors that are entirely absent from the first-order saltation matrix. Consequently, the higher-order expressions for flight time and the associated trajectory deviation map (TDM) provide significantly more accurate predictions of the perturbed trajectory evolution, particularly in regimes characterized by near-grazing impacts.

\begin{figure}[tbh]
	\centering
	\includegraphics[scale = 0.55]{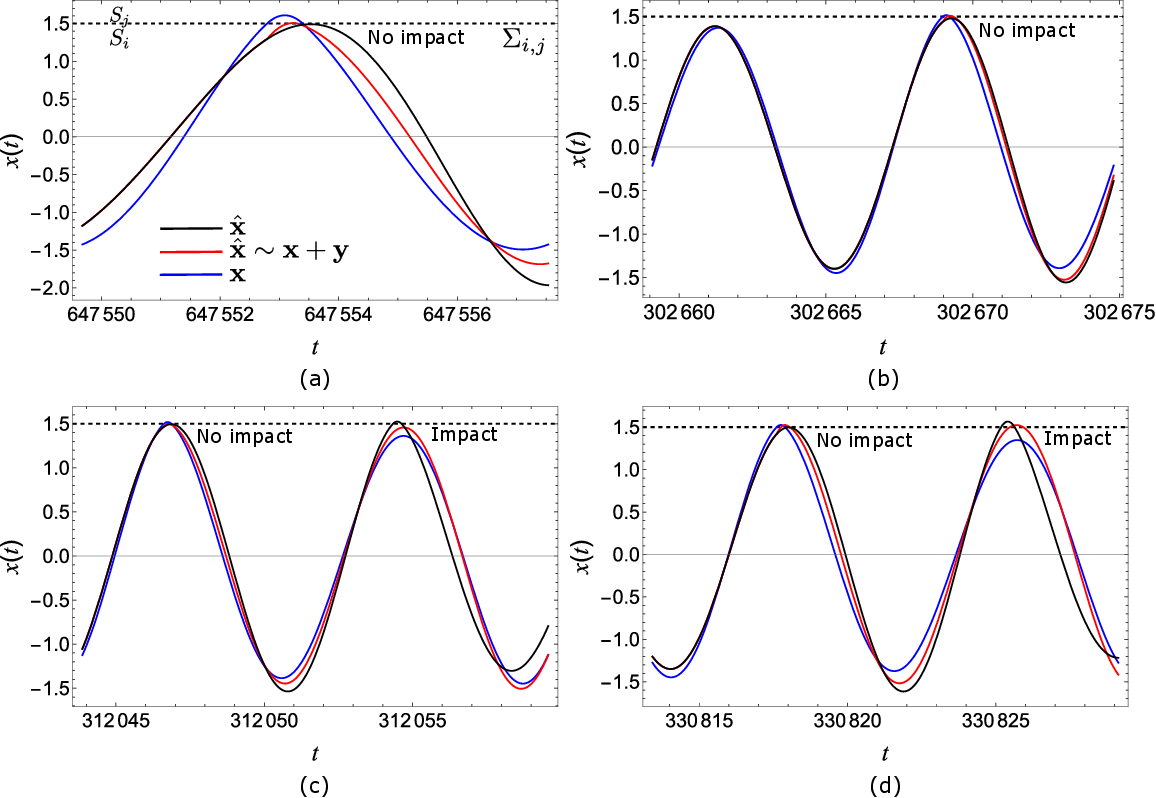}
	\caption{Time series of the displacement \( x(t) \) for (a) a P1 orbit with \( f = 0.7825\,\mathrm{N} \), and P2 orbits for (b) \( f = 0.564\,\mathrm{N} \), (c) \( f = 0.566\,\mathrm{N} \), and (d) \( f = 0.57\,\mathrm{N} \). The periodic orbit \( \mathbf{x} \) and the perturbed orbit \( \mathbf{x} + \mathbf{y} \) are shown in blue and red, respectively. The black curve denotes the true evolution of the perturbed trajectory. In each case, the black curve does not exhibit impact with the discontinuity boundary, in agreement with the prediction from the higher-order flight time \( \delta_+ \in \mathbb{C} \).}
	\label{fig 5}
\end{figure}
Incorrect predictions of impact events can lead to significant discrepancies in trajectory evolution. This is illustrated using the four impacting orbits previously discussed in Fig.~\ref{fig 4}. In Fig.~\ref{fig 5}(a), the time series of a P1 limit cycle \( \mathbf{x} \) (blue curve) and its perturbed counterpart \( \mathbf{x} + \mathbf{y} \) (red curve) are shown over one forcing period for \( f = 0.7825 \, \mathrm{N} \). The first-order approximation predicts an impact of the perturbed trajectory with \( \Sigma_{i,j} \), while the higher-order correction indicates no such event. The true evolution, represented by the black curve, confirms the absence of impact and remains within the region \( S_i \), diverging from the red trajectory and resulting in qualitatively different dynamics.

Similar behavior is observed for the P2 limit cycles shown in Figs.~\ref{fig 5}(b)--(d), corresponding to the cases \( f = 0.564 \), \( 0.566 \), and \( 0.57 \, \mathrm{N} \), respectively, as introduced in Fig.~\ref{fig 4}. In Fig.~\ref{fig 5}(b), the first-order analysis again incorrectly predicts an impact, whereas the higher-order correction correctly anticipates that the perturbed trajectory remains in \( S_i \), a result verified by the true trajectory (black curve). In Fig.~\ref{fig 5}(c), the black curve initially avoids \( \Sigma_{i,j} \), as predicted by the higher-order approximation, but subsequently returns to intersect it---an event not captured by the first-order model, which predicts only a single impact and no return. Similarly, in Fig.~\ref{fig 5}(d), the true perturbed trajectory initially avoids the boundary, consistent with \( \delta_+ \in \mathbb{C} \), but later returns and impacts \( \Sigma_{i,j} \). In contrast, the red trajectory (first-order approximation) impacts early and diverges from the primary orbit, eventually crossing the boundary when the primary does not, rendering the first-order TDM inapplicable.

These results highlight the limitations of first-order flight time and TDM approximations, particularly near grazing or low-velocity impacts. In contrast, the higher-order correction \( \delta_+ \) provides a more accurate prediction of impact occurrence, faithfully capturing the behavior of perturbed trajectories in the vicinity of discontinuity boundaries.

\begin{figure}[tbh]
	\centering
	\includegraphics[scale = 0.785]{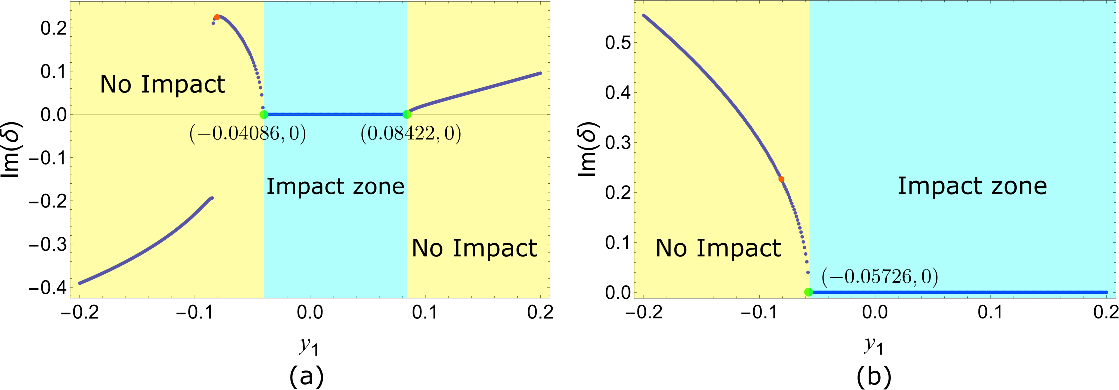}
	\caption{Imaginary part of higher-order flight time $\delta_+$ vs. first component of perturbation $y_1$ for (a) fixed $||\mathbf{y}_-|| = 0.0842268$ and (b) fixed $y_2 = 0.02299$. The higher-order $\delta_+$ predicts that impacts occur in (a) when $-0.0408646 \leq y_1 \leq 0.0842249$ (green points) and (b) when $y_1 = -0.05726$ (green point). The red point corresponds to the value of $\delta_+$ where no impact occurs for $\mathbf{y}_- = \{-0.081026, 0.02299\}$ as verified in Figs. \ref{fig 4}(h) and \ref{fig 5}(d).}
	\label{fig 6}
\end{figure}
Next, we examine the maximum permissible magnitude of perturbations that result in impacts with the discontinuity boundary. Figure~\ref{fig 6}(a) displays the imaginary component of \( \delta_+ \) as the first component of the perturbation vector \( \mathbf{y}_- \), denoted \( y_1 \), is varied, while maintaining a fixed norm \( \| \mathbf{y}_- \| \) during an impact event at \( t_i = 330818\,\mathrm{s} \). The higher-order analysis reveals that impacts occur only when the perturbation magnitude remains within a specific threshold—that is, when \( y_1 \) lies within a bounded range. This permissible interval can be derived analytically by imposing the condition \( B^2 - 4AC \geq 0 \) in Eq.~\eqref{eq 12}, ensuring a real-valued solution for \( \delta_+ \). For the system defined in Eq.~\eqref{eq 20}, this condition becomes:

\begin{equation}\label{eq 23}
	(2m(v + y_2))^2 - 8m(f \cos{\omega t} - k_1 \sigma - c_1 v)y_1 \geq 0,
\end{equation}

where the impact state is \( \mathbf{x}_i = \{\sigma, v\} = \{1.5, -0.349336\} \) at \( t_i = 330818\,\mathrm{s} \), and the perturbed state is \( \mathbf{y}_- = \{y_1, y_2\} \), for the case of \( f = 0.57\,\mathrm{N} \). 

In Fig.~\ref{fig 6}(a), the imaginary part of \( \delta_+ \) is plotted as a function of the first perturbation component \( y_1 \), with the norm \( \|\mathbf{y}_-\| \) held constant during impact. The higher-order results indicate that impacts are possible only when \( y_1 \) lies within a bounded interval. This range, computed analytically using Eq.~\eqref{eq 23}, is given by \( -0.04086 \leq y_1 \leq 0.08422 \). 

Similarly, Fig.~\ref{fig 6}(b) shows the variation of the imaginary part of \( \delta_+ \) with respect to \( y_1 \), keeping \( y_2 \) fixed at \( y_2 = 0.02299 \). The higher-order correction yields an imaginary \( \delta_+ \), indicating that no impact occurs when \( y_1 \leq -0.05726 \), in agreement with the analytical prediction from Eq.~\eqref{eq 23}. 

The selected values of \( \mathbf{x}_i \), \( t_i \), and \( \mathbf{y}_- \) correspond to the scenarios depicted in Figs.~\ref{fig 3}, \ref{fig 4}(g)--(h), and \ref{fig 5}(d), where the true perturbed trajectory exhibits no impact. Therefore, the higher-order solution for \( \delta_+ \) provides a critical threshold that delineates the range of perturbations leading to impact. This also defines the regime where the first-order discontinuity map fails to accurately capture the system dynamics.

\begin{figure}[tbh]
	\centering
	\includegraphics[scale = 0.79]{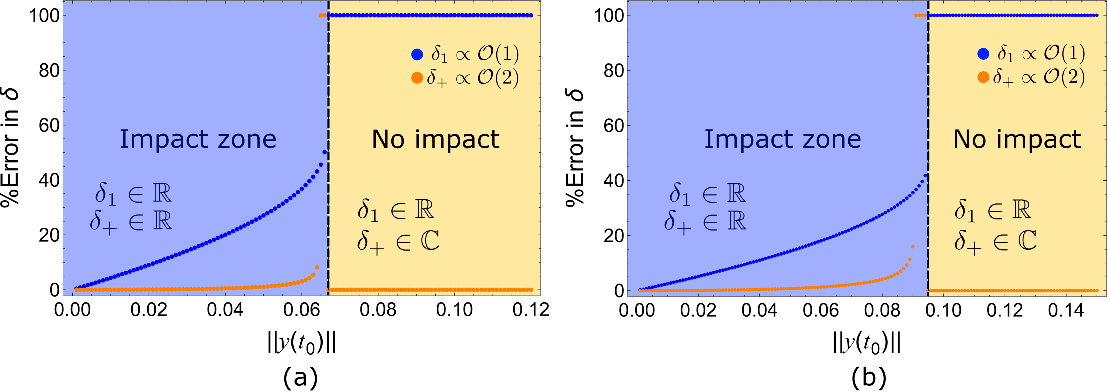}
	\caption{Percent error in the flight time estimate between the first-order approximation \( \delta_1 \) and the higher-order correction \( \delta_+ \), plotted against the initial separation \( \|\mathbf{y}(t_0)\| \) between two nearby orbits. The blue and yellow regions, separated by a black dashed line, represent the zones where impact and no impact occur, respectively, as verified by the exact solution.}
	\label{fig 7}
\end{figure}
Fig.~\ref{fig 7}(a) and (b) compares the percent error in the flight time estimates \( \delta \), as predicted by the first-order approximation \( \delta_1 \) and the higher-order correction \( \delta_+ \) (computed using Eqs.~\eqref{eq 22}), for two P2 orbits corresponding to \( f = 0.564\,\mathrm{N} \) and \( f = 0.57\,\mathrm{N} \), respectively. In Fig.~\ref{fig 7}(a), the primary P2 orbit with \( f = 0.564\,\mathrm{N} \) impacts the discontinuity boundary at some instant \( t_i \). A nearby perturbed orbit with initial perturbation norm \( \|\mathbf{y}(t_0)\| \) reaches the boundary at \( t_i + \delta \). The estimated flight times are compared against the true solution to determine whether the perturbed trajectory actually reaches the boundary.

The blue and yellow regions in Figs.~\ref{fig 7}(a) and (b) represent the impact and no-impact zones, respectively, based on whether the perturbed trajectory intersects the discontinuity boundary. In the impact region, both \( \delta_1 \) and \( \delta_+ \) are real, although the higher-order correction provides more accurate mapping of the perturbed state. In the no-impact region, \( \delta_1 \) remains real but incorrectly predicts an impact, while \( \delta_+ \) becomes imaginary, indicating that no physical root exists and thus no impact occurs. This demonstrates the ability of the higher-order correction to correctly predict impact behavior.

Similarly, Fig.~\ref{fig 7}(b) shows the percent error in the flight time estimate for the case \( f = 0.57\,\mathrm{N} \), further confirming the accuracy of the higher-order method over the first-order approximation.

\begin{figure}[tbh]
	\centering
	\includegraphics[scale = 0.55]{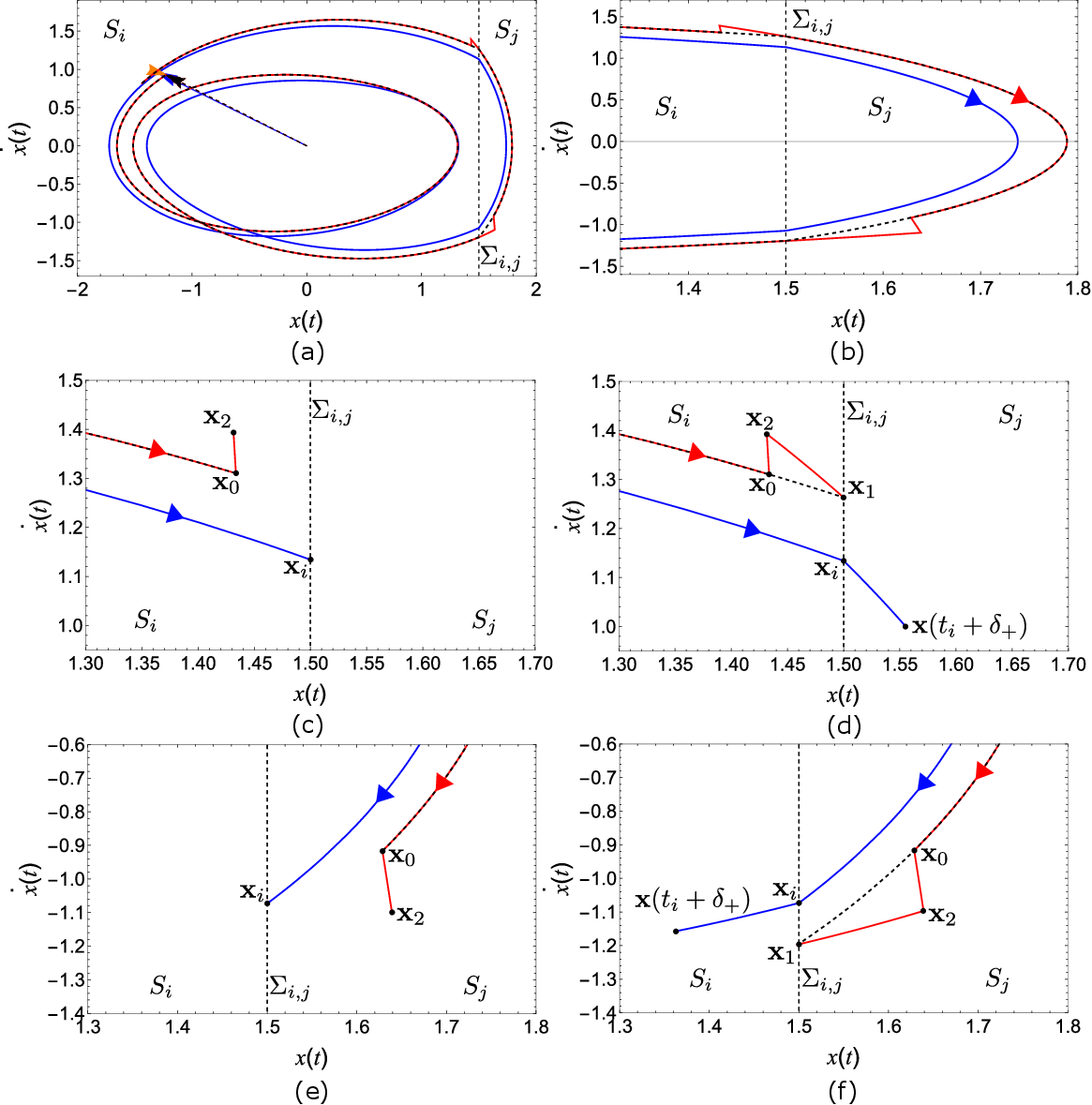}
	\caption{(a) Phase-portrait of a P2 orbit (blue curve) for $f = 0.705$N and a perturbed orbit (red curve) undergoing TDM. (b) Zoomed phase portraits of (a) executing higher-order TDM near the discontinuity barrier. The TDM maps $\mathbf{x}_0$ to $\mathbf{x}_2$ when primary orbit reaches $\Sigma_{i,j}$ at $t = t_i$ while orbits cross $\Sigma_{i,j}$ from (c) $S_i$ to $S_j$ and (e) $S_j$ to $S_i$. Evolution of orbits post higher-order TDM to reach $\mathbf{x}_1$ on $\Sigma_{i,j}$ after predicted flight time $t = t_i + \delta_+$ while orbits cross $\Sigma_{i,j}$ from (d) $S_i$ to $S_j$ and (f) $S_j$ to $S_i$. The black dashed orbit is the true solution and coincides with the predicted paths.}
	\label{fig 8}
\end{figure}

Next, the application of the TDM is illustrated. Fig.~\ref{fig 8}(a) presents the phase portraits of two closely initiated P2 orbits for \( f = 0.705\,\mathrm{N} \). The blue curve represents the primary P2 orbit, while the red curve shows a perturbed orbit initialized with \( \mathbf{y}(t_0) = -\frac{0.2}{\sqrt{2}}\{1,\,1\} \). For comparison, the exact numerical solution of the perturbed orbit is shown as a black dashed curve.

Fig.~\ref{fig 8}(b) illustrates the application of the TDM near the discontinuity boundary \( \Sigma_{i,j} \), where the perturbed orbit crosses from region \( S_i \) to \( S_j \), and subsequently returns to \( S_i \). In Fig.~\ref{fig 8}(c), the perturbed trajectory reaches the point of impact \( \mathbf{x}_0 \) at time \( t_i \), and is subsequently mapped to a new state \( \mathbf{x}_2 \), as defined by the higher-order TDM:

\begin{equation}\label{eq 24}
\mathbf{y}_+ =
\begin{pmatrix}
y_1 - \dfrac{\delta_+^2}{2m}(k_2 \sigma + c_2 v) \\[10pt]
y_2 + \dfrac{\delta_+}{m} \left[ k_2(\sigma + y_1) + c_2(v + y_2) \right]
+ \dfrac{\delta_+^2}{2m^2} \left[ m k_2 v + k_2(c_1 + c_2)\sigma \right. \\
 \left. + c_2(c_2 v - k_1 \sigma) + c_2 f \cos(\omega t_i) \right]
\end{pmatrix}_{2 \times 1}
\end{equation}

For the system given by Eq.~\eqref{eq 20}, when the orbit crosses the discontinuity boundary $\Sigma_{i,j}$ from region $S_i$ to $S_j$. In Fig.~\ref{fig 8}(d), the mapped state $\mathbf{x}_2$ accurately reaches the point $\mathbf{x}_1 \in \Sigma_{i,j}$, in agreement with the theoretical prediction. This result is confirmed by the exact numerical simulation, shown as black dashed lines, which also intersects $\mathbf{x}_1$ on the discontinuity boundary at the instant $t_i + \delta_+$. Likewise, in Fig.~\ref{fig 8}(e), when the trajectory crosses $\Sigma_{i,j}$ from $S_j$ back to $S_i$, the perturbed state first reaches $\mathbf{x}_0$ and is subsequently mapped to $\mathbf{x}_2$, as defined by the higher-order TDM:
\begin{equation}\label{eq 25}
	\mathbf{y}_+ = \begin{pmatrix}
		y_1 + \dfrac{\delta_+^2}{2m}(k_2 \sigma + c_2 v) \\
		\\
		y_2 - \dfrac{\delta_+}{2m}\left[k_2(\sigma + y_1) + c_2(v + y_2)\right] + \dfrac{\delta_+^2}{2m^2}\left[-m k_2 v + k_2(c_2 - c_1)\sigma \right. \\
		\left. + c_2(k_1 \sigma + c_2 v) - c_2 f \cos(\omega t_i)\right]
	\end{pmatrix}_{2 \times 1}
\end{equation}

In Fig.~\ref{fig 8}(f), the mapped state $\mathbf{x}_2$ reaches the point $\mathbf{x}_1 \in \Sigma_{i,j}$ at the instant $t_i + \delta_+$. This result is in good agreement with the numerically simulated perturbed trajectory, shown as a black dashed line, as both the analytical and numerical solutions reach the same point $\mathbf{x}_1$ on the discontinuity boundary $\Sigma_{i,j}$.

\begin{figure}[tbh]
	\centering
	\includegraphics[scale = 0.76]{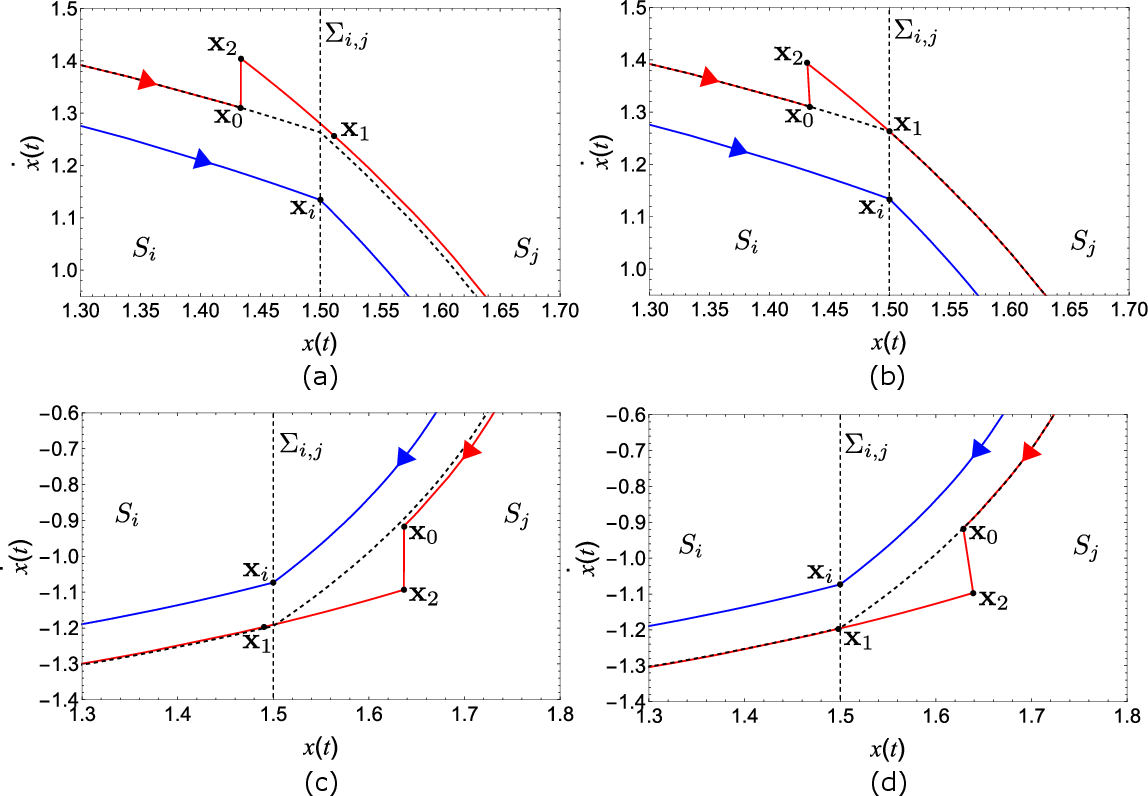}
\caption{Comparison of the TDM from $\mathbf{x}_0$ to $\mathbf{x}_2$ for a perturbed orbit $\mathbf{x} + \mathbf{y}$ in the local neighborhood of the primary orbit $\mathbf{x}$ impacting the discontinuity boundary $\Sigma_{i,j}$ at $\mathbf{x}_i$: (a) first-order TDM when crossing $\Sigma_{i,j}$ from $S_i$ to $S_j$, (b) higher-order TDM for the same transition, (c) first-order TDM when crossing from $S_j$ to $S_i$, and (d) higher-order TDM for the reverse transition. Orbits mapped using the higher-order TDM accurately reach the predicted intersection point $\mathbf{x}_1$ on $\Sigma_{i,j}$, as validated by the true numerical trajectory shown as a black dashed curve.}
\label{fig 9}
\end{figure}
Fig.~\ref{fig 9} compares the performance of the first-order and higher-order transverse discontinuity maps (TDM) for the primary trajectory $\mathbf{x}$ and its perturbed counterpart $\mathbf{x} + \mathbf{y}$, initiated from the perturbation $\mathbf{y}(t_0) = -0.2/\sqrt{2}\{1, 1\}$ with forcing amplitude $f = 0.705$~N. In Fig.~\ref{fig 9}(a), the perturbed state, shown as a red trajectory, begins at $\mathbf{x}_0$ and is mapped to $\mathbf{x}_2$ using the first-order TDM, i.e., the saltation matrix, given by
\begin{equation}\label{eq 26}
	\mathbf{y}_+ = \begin{pmatrix}
		y_1\\
		\\
		\frac{-1}{m v}(k_2 \sigma + c_2 v)y_1 + y_2
	\end{pmatrix}_{2 \times 1}.
\end{equation}

Note that the orbits are evolving from region $S_i$ to $S_j$. The perturbed trajectory reaches the point $\mathbf{x}_1$ after the first-order flight time $t_i + \delta_1$. However, it is evident that $\mathbf{x}_1$ does not lie on the discontinuity boundary $\Sigma_{i,j}$. Moreover, the mapped trajectory fails to coincide with the true perturbed solution (shown as a black dashed line), indicating a lack of accuracy in the first-order approximation.

In contrast, Fig.~\ref{fig 9}(b) shows the mapping of the state $\mathbf{x}_0$ to $\mathbf{x}_2$ using the higher-order TDM. In this case, the trajectory reaches the point $\mathbf{x}_1$ at time $t_i + \delta_+$, which lies precisely on the discontinuity boundary $\Sigma_{i,j}$. The mapped trajectory, computed using higher-order corrections, aligns well with the true solution, demonstrating improved accuracy.

When crossing $\Sigma_{i,j}$ in the reverse direction, from $S_j$ to $S_i$, the perturbed state $\mathbf{x}_0$ in Fig.~\ref{fig 9}(c) is mapped to $\mathbf{x}_2$ using the first-order TDM.

\begin{equation}\label{eq 27}
	\mathbf{y}_+ = \begin{pmatrix}
		y_1\\
		\\
		\frac{1}{m v}(k_2 \sigma + c_2 v)y_1 + y_2    
	\end{pmatrix}_{2 \times 1}
\end{equation}

The resulting trajectory eventually reaches the point $\mathbf{x}_1$. However, this state $\mathbf{x}_1$ overshoots the discontinuity boundary $\Sigma_{i,j}$ due to the influence of the denominator term $\nabla H(\mathbf{x}_i)^T \cdot \mathbf{F}_i(\mathbf{x}_i)$ in the first-order flight time $\delta_1$. 

In contrast, the higher-order TDM provides a more accurate mapping of the initial perturbed state $\mathbf{x}_0$ to $\mathbf{x}_2$, which successfully reaches $\mathbf{x}_1 \in \Sigma_{i,j}$ at the corrected flight time $t_i + \delta_+$, as illustrated in Fig.~\ref{fig 9}(d). The higher-order correction terms in the TDM effectively eliminate the overestimation error observed in the first-order approach.

The following section presents methodologies for conducting stability analysis by incorporating the higher-order flight time estimates and the second-order ($\mathcal{O}(2)$) Transverse Discontinuity Map (TDM) derived in Section~\ref{sec 2}. Approaches for computing Floquet multipliers and Lyapunov exponents are detailed, and their validity is demonstrated through comparison with bifurcation diagrams obtained from both numerical simulations and experimental observations.

\section{Stability analysis}
\label{sec 4}

\begin{figure}[tbh]
	\centering
	\includegraphics[scale = 0.78]{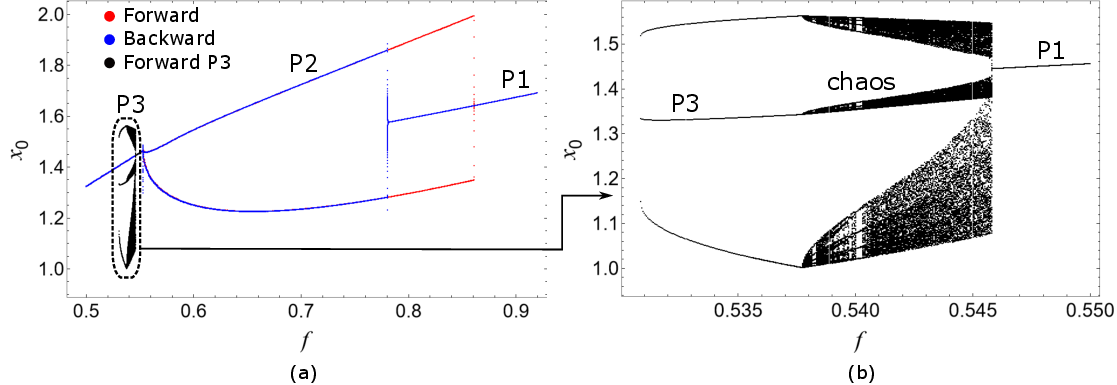}
	\caption{Bifurcation diagrams showing the stroboscopic displacement \(x_0\) as a function of the forcing amplitude \(f\). (a) The red and blue curves correspond to displacement data obtained by incrementally increasing and decreasing \(f\), respectively, over the range \(0.50 \leq f \leq 0.92\)~N. The black curve highlights a period-3 (P3) orbit that coexists in close proximity to the period-1 (P1) orbit. (b) A detailed view illustrating the coexistence of P3 and aperiodic trajectories with the P1 orbit as \(f\) is varied. The presence of multiple coexisting attractors and non-periodic behavior in specific regions indicates rich and complex bifurcation dynamics.}
	\label{fig 10}
\end{figure}
This section illustrates the dynamical behavior of the soft impact oscillator as the external forcing amplitude \(f\) is varied. Figure~\ref{fig 10} presents the bifurcation diagram of the stroboscopically sampled displacement \(x_0\), where sampling is performed once every forcing period \(T = 2\pi/\omega\), following a transient removal phase of \(400T\). In Fig.~\ref{fig 10}(a), the red and blue curves correspond to forward and backward continuation results, respectively, obtained by incrementally increasing and decreasing the value of bifurcation parameter, \(f\), within the range \(0.50 \leq f \leq 0.92\)~N. At each parameter step, the steady-state displacement \(x_0\) is recorded and used as the initial condition for the subsequent value of \(f\).

Upon increasing \(f\), a period-1 (P1) orbit undergoes a bifurcation at \(f = 0.5535\)~N, leading to the emergence of a stable period-2 (P2) orbit. This P2 orbit persists until \(f = 0.861\)~N, at which point it vanishes, and a stable P1 orbit re-emerges and continues up to \(f = 0.92\)~N. In contrast, during the reverse sweep (Fig.~\ref{fig 10}(a), blue curve), the P1 orbit remains stable down to \(f = 0.7815\)~N, below which it disappears and the system transitions to a P2 orbit. This P2 orbit coincides with the forward branch (red curve), indicating the presence of bistability and hysteresis in the system. Depending on the initial conditions, both P1 and P2 orbits can coexist and be observed within the overlapping parameter region.

Additionally, as shown in Fig.~\ref{fig 10}(b), a stable period-3 (P3) orbit (black curve) coexists with the P1 orbit in a narrow range near \(f = 0.53085\)~N to \(0.5377\)~N. To better understand the transitions between these coexisting attractors and to characterize their stability, a Floquet analysis is conducted in the following section, utilizing the higher-order transition discontinuity map (TDM) formalism developed in Sec.~\ref{sec 2}.

\subsection{Floquet multipliers}
This section outlines the methodology for evaluating the global state transition matrix (STM), or monodromy matrix, and its associated Floquet multipliers for Filippov systems. The stability of periodic orbits can be assessed through an eigenvalue analysis of the monodromy matrix. However, in piecewise-smooth (PWS) systems, the monodromy matrix cannot be computed directly using standard matrix exponentials or fundamental solution matrices (FSMs) due to the presence of discontinuity boundaries. These discontinuities necessitate the incorporation of saltation matrices into the STM at each instance of impact. While first-order approaches employ the saltation matrix as defined in Eq.~\eqref{eq 18}, extending these methods to include higher-order corrections is nontrivial. This difficulty arises because the higher-order mapping of the post-impact perturbation $\mathbf{y}_+$ contains nonlinear terms such as $\mathcal{O}(\delta_+^2,\, \delta_+ \mathbf{y}_-,\, \mathbf{y}_- \cdot \mathbf{y}_-)$, which precludes the derivation of a closed-form linear transformation between pre- and post-impact perturbations, as illustrated in Eq.~\eqref{eq 18}. To address this, a numerical procedure is proposed to compute the higher-order saltation matrix required for accurately evaluating the STM in Filippov systems.

Let $\mathbf{x} \in \mathbb{R}^n$ represent the state vector of a Filippov dynamical system of dimension $n$. The evolution of $\mathbf{x}$ and its infinitesimal perturbations $\mathbf{y}$ in the local neighborhood is governed by Eqs.~\eqref{eq 2} and \eqref{eq 3}, respectively. To compute the Lyapunov spectrum, an orthonormal basis of perturbation vectors is initialized on the surface of a hypersphere with radius $r_0$. These perturbation vectors can be compactly represented in matrix form as $\mathbf{Y}$, where each column corresponds to a perturbation vector $\mathbf{y}_i$ for $i = 1, 2, \ldots, n$, i.e.,
\begin{equation}\label{eq 28}
	\mathbf{Y}_{n \times n}(t) = r_0[\mathbf{y}_1, \mathbf{y}_2, \ldots, \mathbf{y}_i, \mathbf{y}_{i + 1}, \ldots, \mathbf{y}_n],
\end{equation}
ensuring that the perturbations span the local tangent space of the phase space of the system at time $t$.

It is evident that the evolution of the matrix-valued function $\mathbf{Y}(t)$ is governed by the Jacobian of the vector field associated with the corresponding subspace $S_i$. Let $\mathbf{Y}_{-, \text{impact}}$ and $\mathbf{Y}_{+, \text{impact}}$ denote the matrices of perturbation vectors immediately before and after impact, respectively, as mapped through the higher-order transverse discontinuity mapping (TDM) defined in Eq.~\eqref{eq 17}. These matrices are expressed as:
\begin{equation} \label{eq 29}
		\mathbf{Y}_{+, \text{impact}} = \begin{bmatrix}
			y^{(1)}_{+, 1} & \ldots & y^{(1)}_{+, n} \\
			\vdots & \ddots & \vdots \\
			y^{(n)}_{+, 1} & \ldots & y^{(n)}_{+, n}
		\end{bmatrix}_{n \times n} \hspace{0.5cm} \text{and} \hspace{0.5cm}
		\mathbf{Y}_{-, \text{impact}} = \begin{bmatrix}
			y^{(1)}_{-, 1} & \ldots & y^{(1)}_{-, n} \\
			\vdots & \ddots & \vdots \\
			y^{(n)}_{-, 1} & \ldots & y^{(n)}_{-, n}
		\end{bmatrix}_{n \times n},
\end{equation}
where $y^j_{-,i}$ and $y^j_{+,i}$ denote the $j^{\text{th}}$ component of the $i^{\text{th}}$ perturbation vector $\mathbf{y}_i$ before and after impact, respectively. A corresponding matrix transformation $\mathbf{S}_2$ can now be defined such that:
\begin{equation}\label{eq 30}
	\mathbf{Y}_{+, \text{impact}} = \mathbf{S}_2 \cdot \mathbf{Y}_{-, \text{impact}}.
\end{equation}
Equivalently, the higher-order saltation matrix $\mathbf{S}_2$ can be expressed as:
\begin{equation}\label{eq 31}
	\mathbf{S}_2 = \mathbf{Y}_{+, \text{impact}} \cdot \mathbf{Y}^{-1}_{-, \text{impact}},
\end{equation}
where $\mathbf{S}_2$ denotes the numerically computed higher-order state transition matrix (STM) at the instant of impact. This matrix captures the perturbation dynamics more accurately than its first-order counterpart due to the inclusion of the improved flight-time estimate $\delta_+$. 

The next objective is to compute the monodromy matrix for a periodic orbit using the higher-order saltation matrix given by Eq.~\eqref{eq 31}. The monodromy matrix is obtained as a product of STMs, including both the continuous-time flow maps and the discrete saltation matrices, composed in the order of their occurrence throughout the orbit.

\begin{figure}[tbh]
	\centering
	\includegraphics[scale = 0.9]{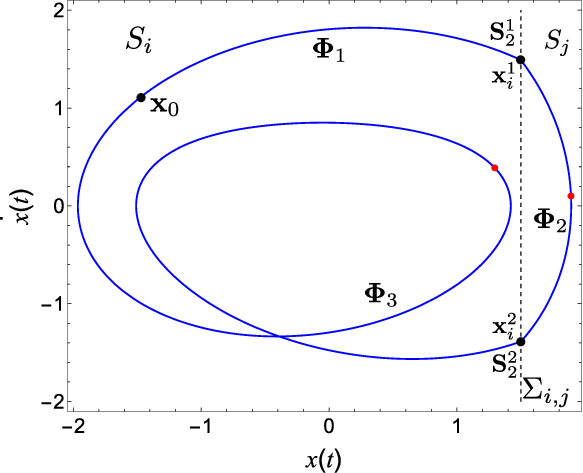}
	\caption{Phase portrait of a period-two (P2) orbit corresponding to a forcing amplitude of $f = 0.80$~N. The red points indicate Poincar\'e intersections. The trajectory undergoes two impacts with the discontinuity boundary $\Sigma_{i,j}$ at the points $\mathbf{x}^1_i$ and $\mathbf{x}^2_i$. The monodromy matrix for this orbit is constructed by sequentially combining the state transition matrices (STMs) $\mathbf{\Phi}_1$, $\mathbf{\Phi}_2$, and $\mathbf{\Phi}_3$ with the associated higher-order saltation matrices $\mathbf{S}^1_2$ and $\mathbf{S}^2_2$ at the respective impact events.}
	\label{fig 11}
\end{figure}
To illustrate the procedure, consider the phase portrait of a period-two (P2) orbit shown in Fig.~\ref{fig 11}, corresponding to a forcing amplitude of $f = 0.80$~N. The orbit is initiated from a state $\mathbf{x}(t_0)$ and undergoes two impacts with the discontinuity boundary $\Sigma_{i,j}$ at times $t^1_i$ and $t^2_i$, reaching impact points $\mathbf{x}^1_i(t^1_i)$ and $\mathbf{x}^2_i(t^2_i)$, respectively, before returning to the initial point after a duration of $t = t_0 + 2T$.

Let $\mathbf{\Phi}_1(t)$, $\mathbf{\Phi}_2(t)$, and $\mathbf{\Phi}_3(t)$ denote the fundamental solution matrices (FSMs), each satisfying $\mathbf{\Phi}_1(0) = \mathbf{\Phi}_2(0) = \mathbf{\Phi}_3(0) = \mathbb{I}_{n \times n}$, corresponding to the flow in regions $S_i$, $S_j$, and $S_i$, respectively. Specifically:
- $\mathbf{\Phi}_1(t)$ governs the flow from $\mathbf{x}(t_0)$ to $\mathbf{x}^1_i$ within $S_i$,
- $\mathbf{\Phi}_2(t)$ governs the flow from $\mathbf{x}^1_i$ to $\mathbf{x}^2_i$ within $S_j$,
- $\mathbf{\Phi}_3(t)$ governs the return flow from $\mathbf{x}^2_i$ to $\mathbf{x}(t_0 + 2T)$ within $S_i$.

Consequently, the monodromy matrix associated with the P2 orbit can be expressed as
\begin{equation} \label{eq 32}
    \mathbf{\Phi} = \mathbf{\Phi}_3 \cdot \mathbf{S}^2_2 \cdot \mathbf{\Phi}_2 \cdot \mathbf{S}^1_2 \cdot \mathbf{\Phi}_1,
\end{equation}
where $\mathbf{S}^1_2$ and $\mathbf{S}^2_2$ denote the higher-order saltation matrices corresponding to the first and second discontinuous transitions, respectively, and $\mathbf{\Phi}_i$ represent the state transition matrices (STMs) governing the smooth flow segments between successive impacts.

More generally, for a periodic orbit of type P$k$, consisting of $k$ impacts with the switching manifold $\Sigma_{i,j}$, the monodromy matrix takes the form
\begin{equation} \label{eq 33}
    \mathbf{\Phi}(kT) = \mathbf{\Phi}_{k + 1} \cdot \prod_{i = 1}^{k} \mathbf{S}^i_2 \cdot \mathbf{\Phi}_i,
\end{equation}
where the product is ordered chronologically from the first to the $k$-th impact. Here, $\mathbf{\Phi}_i$ denotes the STM over the $i^{\text{th}}$ smooth segment, while $\mathbf{S}^i_2$ is the corresponding higher-order saltation matrix applied at the $i^{\text{th}}$ discontinuity.

To numerically compute the monodromy matrix as described in Eq.~\eqref{eq 33}, two fundamental solution matrices, $\mathbf{Y}_1(t)$ and $\mathbf{Y}_2(t)$, are integrated in parallel with the system dynamics. The evolution of $\mathbf{Y}_1(t)$ yields the STMs, while $\mathbf{Y}_2(t)$ captures the trajectory deviation across discontinuities, allowing for the construction of saltation matrices at each impact event.
\begin{align}\label{eq 34}
    \mathbf{Y}_1(0) &= r_0 \mathbb{I}_{n \times n}, \quad \text{for the saltation matrices } \mathbf{S}^i_2, \\
    \mathbf{Y}_2(0) &= \mathbb{I}_{n \times n}, \quad \text{for the state transition matrices (STMs) } \mathbf{\Phi}_i. \nonumber
\end{align}
The numerical procedure for calculating the higher-order saltation matrices, monodromy matrix, and its associated Floquet multipliers, as outlined in Eqs.~\eqref{eq 31}, \eqref{eq 33}, and \eqref{eq 34}, is provided in Appendix~\ref{app a}. This methodology is also applicable to hybrid piecewise-smooth (PWS) systems, such as the impact oscillator \cite{chawla2022stability,chawla2023higher,chawla2024wake,chawla2025higher}.


\begin{figure}[tbh]
	\centering
	\includegraphics[scale = 0.75]{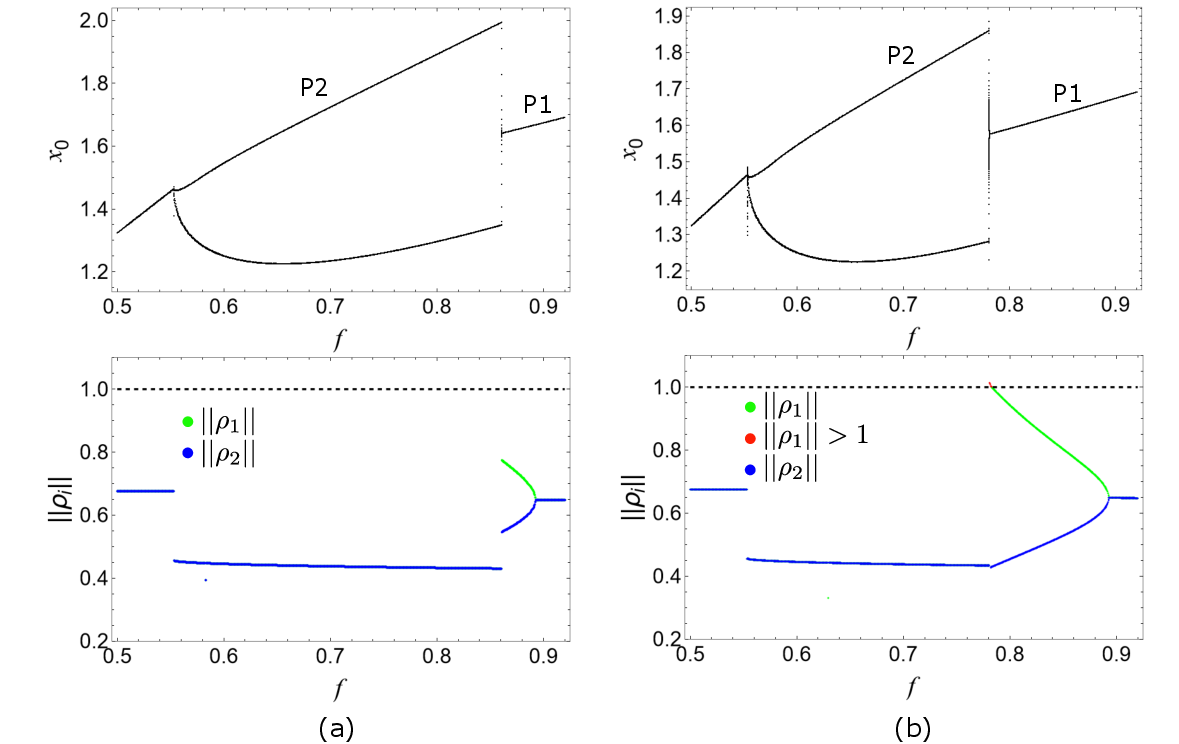}
	\caption{Bifurcation diagram of the stroboscopic displacement \(x_0\) versus the forcing amplitude \(f\) plotted alongside the norms of the Floquet multipliers \(\rho_1\) and \(\rho_2\) while \(f\) is (a) increasing (forward bifurcation) and (b) decreasing (backward bifurcation) in the range \(0.50 \leq f \leq 0.92 \,~\text{N}\). The black dashed lines indicate the unit norm for the Floquet multipliers.}
	\label{fig 12}
\end{figure}
Figs.~\ref{fig 12}(a) and \ref{fig 12}(b) compare the norms of the Floquet multipliers with the forward and backward bifurcation diagrams. In Fig.~\ref{fig 12}(a), the Floquet multipliers remain continuous until a discontinuous change occurs at \( f = 0.5535 \, \text{N} \), where a new stable P2 orbit is formed. As \( f \) is further increased to \( f = 0.861 \, \text{N} \), the norms of the multipliers decrease, reaching values less than unity. At this point, the existing P2 orbit disappears, and a stable P1 orbit emerges. This observation is more pronounced when the Floquet multipliers are evaluated during the decrease in \( f \), as shown in Fig.~\ref{fig 12}(b). Initially, the norms of the multipliers are less than unity, indicating the stability of the P1 orbit. The norms remain unchanged until \( f = 0.8925 \, \text{N} \), after which, upon further decreasing \( f \), one of the Floquet multipliers (depicted by the green curve) becomes greater than unity at \( f = 0.7805 \, \text{N} \). This marks the destabilization of the P1 orbit. Upon further reduction of \( f \), the norms become equal and undergo a discontinuous jump to a value less than unity, indicating the birth of a stable P2 orbit.

\begin{figure}[tbh]
	\centering
	\includegraphics[scale = 0.75]{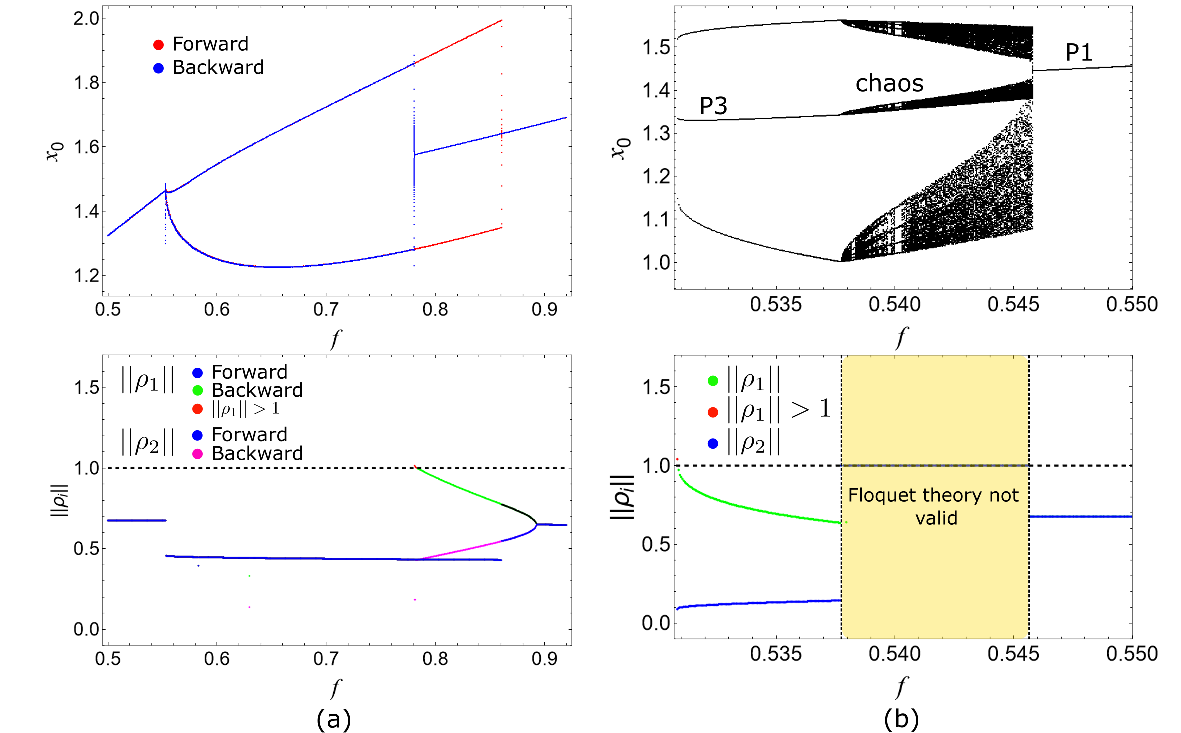}
	\caption{Bifurcation diagram of the stroboscopic displacement \(x_0\) versus the forcing amplitude \(f\) plotted against the norm of Floquet multipliers \(\rho_1\) and \(\rho_2\), where \(f\) is (a) increasing and decreasing within the range \(0.50 \leq f \leq 0.92 \,~\text{N}\), and (b) increasing when P3 orbits coexist with P1 orbits. The yellow region in panel (b) corresponds to aperiodic orbits, where Floquet theory is not applicable.}
	\label{fig 13}
\end{figure}
The overlap of Floquet multiplier values as \( f \) is increased and decreased is shown in Fig.~\ref{fig 13}(a), where the Floquet multipliers are color-coded to highlight the \( f \)-ranges for which the P1 and P2 orbits are stable. Fig.~\ref{fig 13}(b) presents a comparison of the Floquet multiplier norms with bifurcation results for P3 limit cycles near \( f = 0.53085 \, \text{N} \), as previously shown in Fig.~\ref{fig 10}(b). The yellow region in Fig.~\ref{fig 13}(b) corresponds to the range of \( f \) where aperiodic solutions were obtained, rendering the Floquet theory inapplicable. However, within the parameter range where P3 orbits were observed, the norm \( ||\rho_1|| \) approaches and exceeds unity at \( f = 0.53085 \, \text{N} \), indicating that the P3 orbit should become unstable, a result that is confirmed by the bifurcation diagram.

\begin{figure}[tbh]
	\centering
	\includegraphics[scale = 1.2]{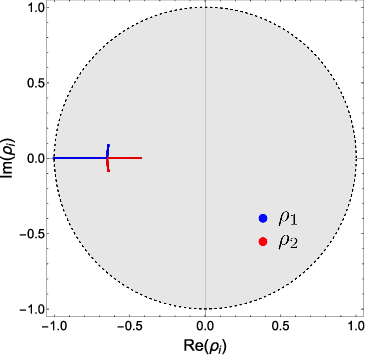}
	\caption{Argand diagram depicting the Floquet multipliers \(\rho_1\) and \(\rho_2\) in the complex plane for the range \(0.7820 \leq f \leq 0.92\) N. The dashed black circle represents the stable region where the Floquet multipliers have unit norm. The P1 orbits, as observed in Figs.~\ref{fig 12}(b) and \ref{fig 13}(a), become unstable at \(f = 0.7830\)~N, where \(\rho_1 = -0.999796\), indicating the onset of instability through a period-doubling bifurcation.}
	\label{fig 14}
\end{figure}
Fig.~\ref{fig 14} presents the Floquet multipliers \(\rho_1\) and \(\rho_2\) in the complex plane for the forcing amplitude range \(0.7820 \leq f \leq 0.92\) N. At the higher end of the range, specifically at \(f = 0.92\)~N, the magnitudes of the multipliers remain within the unit circle, indicating stability. As the forcing amplitude \(f\) decreases, the norms of the eigenvalues increase. At \(f = 0.7830\)~N, the multiplier \(\rho_1\) reaches a value of \(-0.999796\), signaling the imminent destabilization of the P1 orbit. This transition occurs at \(f = 0.7825\)~N, where the orbit becomes unstable through a period-doubling bifurcation. The corresponding instability is further corroborated by the bifurcation diagrams presented in Figs.~\ref{fig 12}(b) and \ref{fig 13}(a).

It is important to note that Floquet theory is applicable solely to periodic solutions. In the following section, we present the calculation of the largest Lyapunov exponent using the higher-order flight time \(\delta_+\) and the TDM, providing a more comprehensive stability analysis of the system.

\subsection{Lyapunov exponents}
Lyapunov exponents quantify the average exponential rates at which two nearby trajectories diverge in phase space, thereby characterizing the stability of a dynamical system. In smooth dynamical systems, the evolution of perturbations is governed by the Jacobian matrix of the underlying system, as described in Eqs.~\eqref{eq 3}. However, for Filippov systems, the Jacobian is discontinuous across the switching manifold, rendering conventional methods inapplicable during impacts. This challenge is addressed by employing the higher-order transverse discontinuity mapping (TDM), as defined in Eqs.~\eqref{eq 22}, \eqref{eq 24}, and \eqref{eq 25}, which accurately captures the evolution of perturbations during discontinuous events.

By integrating the higher-order TDM, one can compute the complete Lyapunov spectrum for a piecewise-smooth (PWS) system of dimension $n$, where each exponent $\lambda_i$ corresponds to a distinct rate of divergence. The Lyapunov exponents are evaluated as time-averaged logarithmic growth rates of infinitesimal perturbations, given by
\begin{equation}\label{eq 35}
	\lambda_i = \frac{1}{\tau N} \sum_{j = 1}^{N} \log\left(\frac{r_i}{r_0}\right),
\end{equation}
Here, $r_i$ denotes the norm of the $i^{\text{th}}$ perturbation vector at each time interval $\tau$, with the initial norm set to $r_0$. These norms are measured stroboscopically over $N$-time steps. To compute the Lyapunov spectrum $\lambda_i$, an ensemble of $n$ orthogonal perturbation vectors is initialized on a hypersphere of radius $r_0$, ensuring mutual orthogonality. The perturbations are then evolved for a time interval $\tau$, after which their norms are recorded.

Orthogonality is preserved throughout the procedure using QR decomposition via the Gram–Schmidt orthonormalization process. This method yields the growth rates $r_i$ corresponding to all $n$ perturbation vectors, aligned with the local eigendirections of the Jacobian matrix. After each time interval $\tau$, the perturbation vectors are rescaled to norm $r_0$, and the process is repeated. The use of QR decomposition ensures that the dominant direction associated with the largest Lyapunov exponent remains unaltered during reorthogonalization.

Since the analysis assumes the validity of local linearization, the initial radius is chosen to be small, $r_0 = 10^{-6}$. For accurate estimation, each Lyapunov exponent $\lambda_i$ is computed using Eq.~\eqref{eq 35}, averaged over 1500 iterations. The complete methodology for computing the Lyapunov spectrum, including the incorporation of the higher-order TDM during impacts, is detailed in Appendix~\ref{app b}.

\begin{figure}[tbh]
	\centering
	\includegraphics[scale = 0.75]{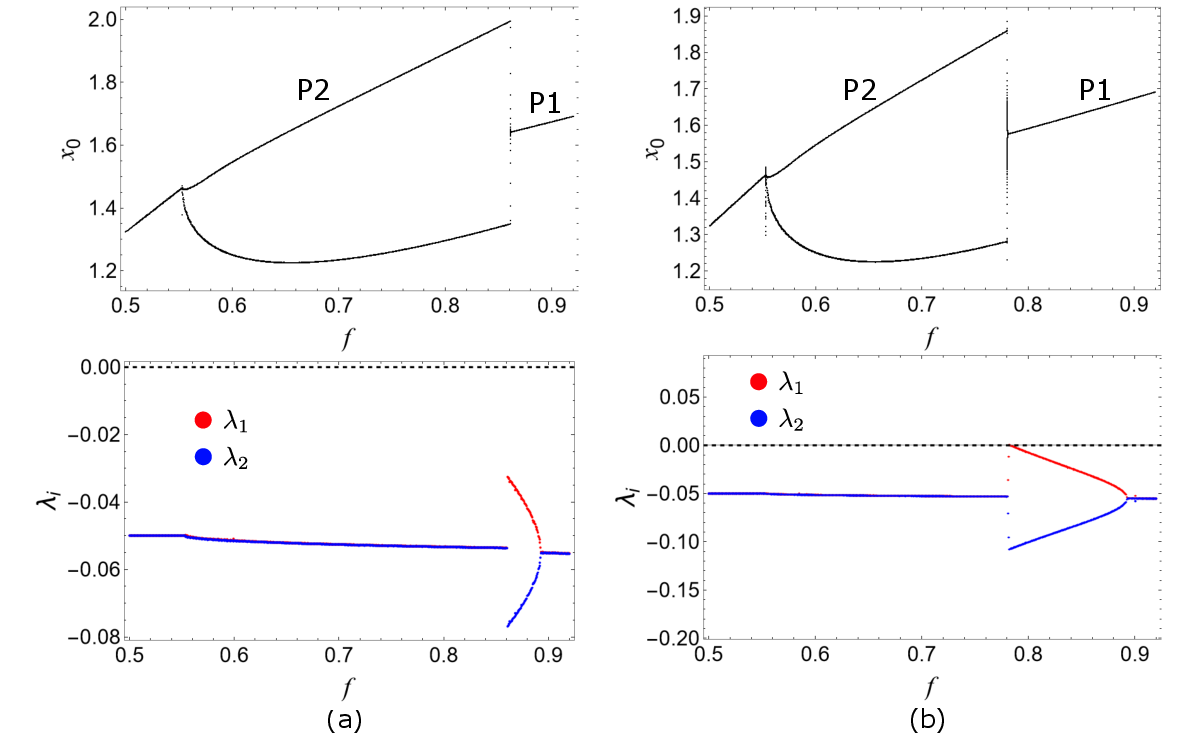}
	\caption{Comparison of the stroboscopic displacement $x_0$ and Lyapunov exponents $\lambda_1$ and $\lambda_2$ with respect to the forcing amplitude $f$ during (a) forward and (b) backward parameter sweeps in the range $0.50 \leq f \leq 0.92$~N.}
	\label{fig 15}
\end{figure}
Figure~\ref{fig 15}(a) and (b) compare the forward and backward bifurcation diagrams of the stroboscopically sampled displacement $x_0$ with the corresponding Lyapunov spectrum computed using the higher-order transverse discontinuity mapping (TDM). In the forward sweep shown in Fig.~\ref{fig 15}(a), both Lyapunov exponents remain negative even after a discontinuous transition at $f = 0.861$~N, where the existing period-2 (P2) orbit vanishes and a new period-1 (P1) orbit emerges. 

In contrast, the backward bifurcation diagram in Fig.~\ref{fig 15}(b) provides a different dynamics. Initially, the Lyapunov exponents are negative, consistent with the stable P1 orbit. As the forcing amplitude $f$ is decreased, the largest Lyapunov exponent $\lambda_1$ increases and becomes positive at $f = 0.7825$~N, signaling the onset of instability in the P1 orbit. This transition corresponds to the emergence of a new stable P2 orbit, consistent with the observed bifurcation behavior.

The following section presents experimental validation of the discontinuity-induced bifurcations (DIBs) observed in Filippov systems, specifically demonstrating the existence of period-1 (P1), period-2 (P2), and period-3 (P3) orbits in the soft-impact oscillator under parametric variation.

\section{Equivalent electronic circuit of an impact oscillator having a pre-stressed barrier} \label{sec 5}

\begin{figure}[tbh]
	\centering
	\includegraphics[scale = 0.8]{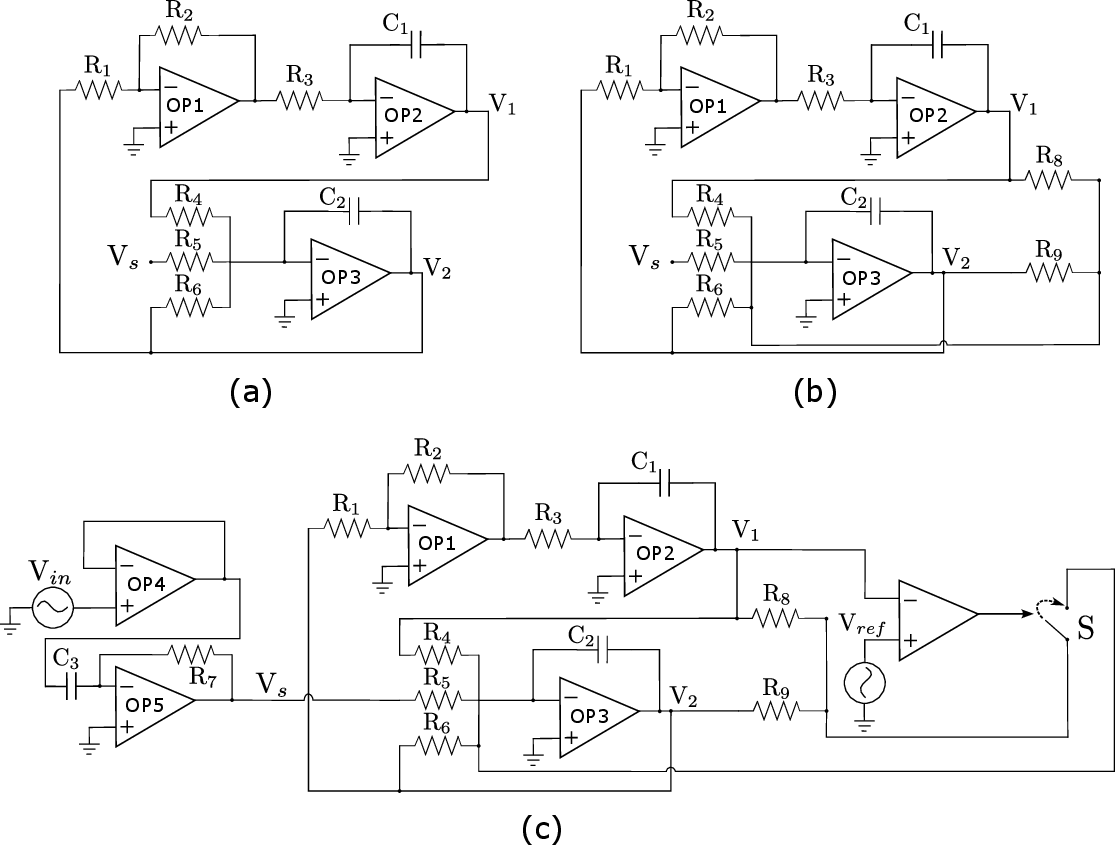}
	\caption{Circuit diagram of equivalent soft impact oscillator (a) before and (b) after impact, and (c) implementation using a comparator IC.}
	\label{fig 16}
\end{figure}
To validate the numerical predictions, we developed an inductorless electronic circuit emulating the soft-impact oscillator with a pre-stressed barrier. The circuit features an operational amplifier (Op-amp), designated OP5, which integrates an input signal from a buffer amplifier OP4. This input consists of a sinusoidal signal with amplitude ($V_{\rm in}$) and a linear frequency ($f_{\rm in}$). This configuration mirrors the external forcing term ($ f\cos {\omega t}$) in the mechanical system, as described by Eq.~(\ref{eq 20}). Additionally, Op-amp OP3 generates the state variable ($\dot{x}$), corresponding to the first derivative of the position, as presented in Eq.~(\ref{eq 20}). A secondary set of Op-amps integrates the output from OP3, thus replicating two first-order differential equations. The Op-amps OP2 and OP3 outputs are designated as the state variables ($V_1$) and ($V_2$), respectively. This method is relatively straightforward and offers advantages over comparable approaches that rely on passive components such as inductors, capacitors, and resistors to achieve similar outcomes. The proposed technique effectively captures discontinuity-induced bifurcations in piecewise linear differential equations, characteristic of mechanical systems, across a broad spectrum of stiffness ratios.

The circuit diagrams in Figs.~\ref{fig 16}(a) and (b) replicate the equations of motion for the soft-impact oscillator, as given by Eq.~(\ref{eq 20}), both before and after impact. The output voltages ($V_1$ and $V_2$) of OP2 and OP3 correspond to the position \(x\) and the velocity \(\dot{x}\) in Eq.~(\ref{eq 20}), respectively. A voltage comparator governs the switching between the two equations. When \(V_1\) is less than \(V_{\rm ref}\), the output of the comparator will go high state, which makes the switch {\em ON}, and a resistive branch is added to the inverting input of OP3. When \(V_1\) is greater than \(V_{\rm ref}\), the switch will be {\em OFF}, and there will be no additional resistive branch. The complete circuit diagram, including the voltage comparator, is shown in Fig.~\ref{fig 16}(c). The governing equations of the circuit depicted in Fig.~\ref{fig 16}(c) are given by:
\begin{align} \label{eq 36}
	V_1 &= \frac{1}{C_1 R_3}\int \Big(\frac{-R_2}{R_1} V_2\Big) dt\\
    V_2 &= \begin{cases} \nonumber
		-\frac{1}{C_2} \int \Big(\frac{V_s}{R_5} + \frac{V_1}{R_4} + \frac{V_2}{R_6}\Big) dt, \qquad V_{\text{ref}} < 1\\
		-\frac{1}{C_2} \int \Big(\frac{V_s}{R_5} + \frac{V_1}{R_4} + \frac{V_2}{R_6}\Big) - \frac{1}{C_2} \int \Big(\frac{V_1}{R_8} + \frac{V_2}{R_9}\Big) dt, \qquad V_{\text{ref}} \geq 1
	\end{cases}
\end{align}

Eq.~(\ref{eq 20}) must be non-dimensionalized to facilitate direct, one-to-one comparison between the mechanical system and the proposed analog electronic circuits. To achieve this, we introduce the following dimensionless quantities:
\begin{equation}
\tau = \omega_1 t, \quad x_d = \frac{x}{d}, \quad \xi_1 = \frac{c_1}{2 m \omega_1}, \quad \xi_2 = \frac{c_2}{2 m \omega_1}, \quad f_0 = \frac{f}{m \omega_1^2 d}.
\end{equation}

\begin{align} \label{eq 37}
	x^{\prime\prime}_d =
	\begin{cases}
		f_0 \cos{\omega_0 \tau} - x_d  - 2\xi_1 x_d^\prime, \quad & x_d < 1, \\
		f_0 \cos{\omega_0 \tau} - (1 + \beta)x_d - 2(\xi_1 + \xi_2) x^\prime_d, \quad & x_d \geq 1.
	\end{cases}
\end{align}
where the characteristic angular frequency is defined as \(\omega_1^2 = k_1/m\), the normalized driving frequency is given by \(\omega_0 = \omega/\omega_1\), and the stiffness ratio is expressed as \(\beta = k_2/k_1\). The prime notation denotes differentiation with respect to the dimensionless time variable \(\tau\).  
The input voltage to the circuit is externally supplied via a waveform generator and is given by  
\[
V_{in} = A_{in} \sin{(2\pi f_{in} t)}.
\]
The output voltage of operational amplifier 5 (OP5) is  
\[
V_s = -C_3 R_7 \frac{dV_{in}}{dt} = -2 \pi f_{in} C_3 R_7 A_{in} \cos{(2\pi f_{in} t)}.
\]
By selecting the circuit parameters as follows:  
\[
R_1 = R_2 = 7.0~\text{k}\Omega, \quad R_3 = R_4 = R_5 = R_7 = R_8 = R = 10.0~\text{k}\Omega,
\]
\[
R_6 = R_9 = 10R = 100~\text{k}\Omega, \quad C_1 = C_2 = C_3 = C \approx 9.38~\text{nF},
\]

The equivalent circuit equation corresponding to Eq.~\eqref{eq 37} is obtained as:
\begin{align} \label{eq 38}
    V_1 &= \int V_2 \, d\tilde{\tau}, \\
    V_2 &= 
    \begin{cases} 
        \int \Big( \tilde{A} \cos{\tilde{\omega} \tilde{\tau}} - V_1 - 0.1 V_2 \Big) \, d\tilde{\tau}, \quad & V_{\text{ref}} < 1, \\[8pt]
        \int \Big( \tilde{A} \cos{\tilde{\omega} \tilde{\tau}} - 2 V_1 - 0.2 V_2 \Big) \, d\tilde{\tau}, \quad & V_{\text{ref}} \geq 1.
    \end{cases}
\end{align}
where the dimensionless time variable is defined as  
\[
\tilde{\tau} = \frac{t}{C R},
\]
while the normalized amplitude and frequency are given by  
\[
\tilde{A} = 2\pi f_{\text{in}} A_{\text{in}} C R, \quad \text{and} \quad \tilde{\omega} = \omega_0 = 2\pi f_{\text{in}} C R.
\]
Consequently, Eqs.~\eqref{eq 37} and \eqref{eq 38} become equivalent when \(\beta = 1\) and \(\xi_1 = \xi_2 = 0.05\).
For direct experimental validation of the numerical results from Sec.~\ref{sec 4}, the external sinusoidal frequency is set as  
\[
f_{\text{in}} = \frac{\omega_0}{2\pi C R} = 1358.85~\text{Hz},
\]
ensuring that \(\tilde{\omega} = \omega_0 = \omega = 0.8\). Given that the force \(f\) varies within the range \(0.5 \leq f \leq 0.92\) N, the corresponding input voltage amplitude \(A_{\text{in}}\) must satisfy  
\[
0.416 \leq A_{\text{in}} \leq 0.766~\text{V},
\]
ensuring \(\tilde{A} = f_0\). The reference voltage \(V_{\text{ref}}\), which defines the discontinuity boundary, is set to \(V_{\text{ref}} = 1.0\) V, with a comparator IC facilitating the switching mechanism. The circuit diagram, including IC pin configurations, experimental setup, and methodology, is provided in Appendix~C.

\begin{figure}[h!]
	\centering
	\includegraphics[scale = 0.6]{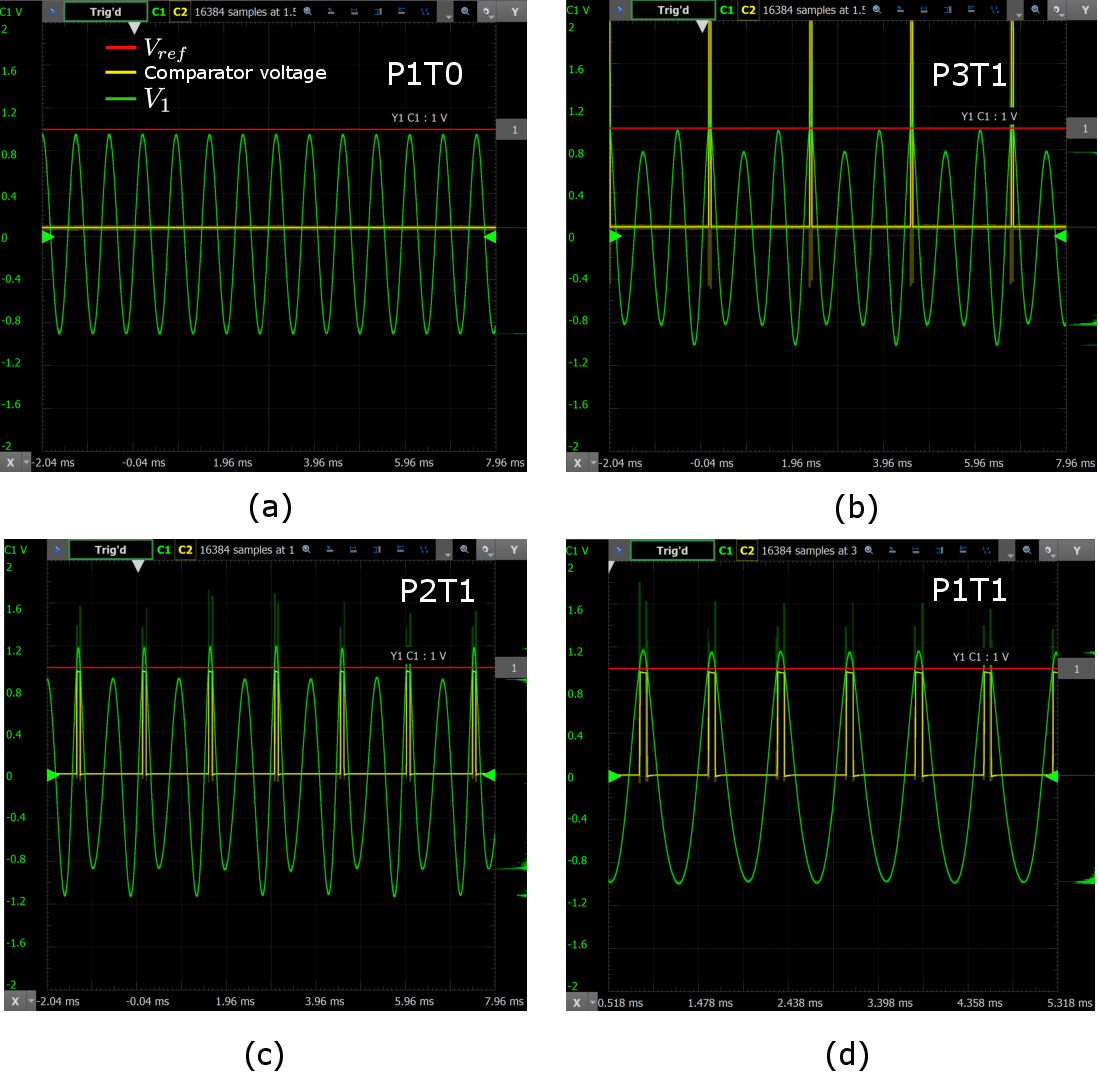}
	\caption{Time series of \(V_1\) illustrating the observed orbits for different input voltage amplitudes:  
(a) P1T0 at \(A_{\text{in}} = 450\) mV,  
(b) P3T1 at \(A_{\text{in}} = 465\) mV,  
(c) P2T1 at \(A_{\text{in}} = 600\) mV, and  
(d) P1T1 at \(A_{\text{in}} = 750\) mV.  
The red and yellow curves represent the reference voltage \(V_{\text{ref}} = 1.0\) V and the comparator output voltage, respectively.}
	\label{fig 17}
\end{figure}
Fig.~\ref{fig 17} presents the measured voltage signals \(V_1\), \(V_{\text{ref}}\), and the comparator output voltage. The time series in Fig.~\ref{fig 17}(a) corresponds to an input amplitude of \(A_{\text{in}} = 450\) mV, exhibiting a P1T0 limit cycle, as \(V_1\) remains below \(V_{\text{ref}}\), keeping the comparator in the {\em OFF} state. As \(V_1\) exceeds \(V_{\text{ref}}\), the comparator IC transitions to the {\em ON} state, resulting in the emergence of distinct limit cycles. Specifically, the P3T1, P2T1, and P1T1 limit cycles are observed in Figs.~\ref{fig 17}(b), \ref{fig 17}(c), and \ref{fig 17}(d) for input amplitudes of \(A_{\text{in}} = 465\)~mV, \(600\)~mV, and \(750\)~mV, respectively.

\begin{figure}[h!]
	\centering
	\includegraphics[scale = 0.8]{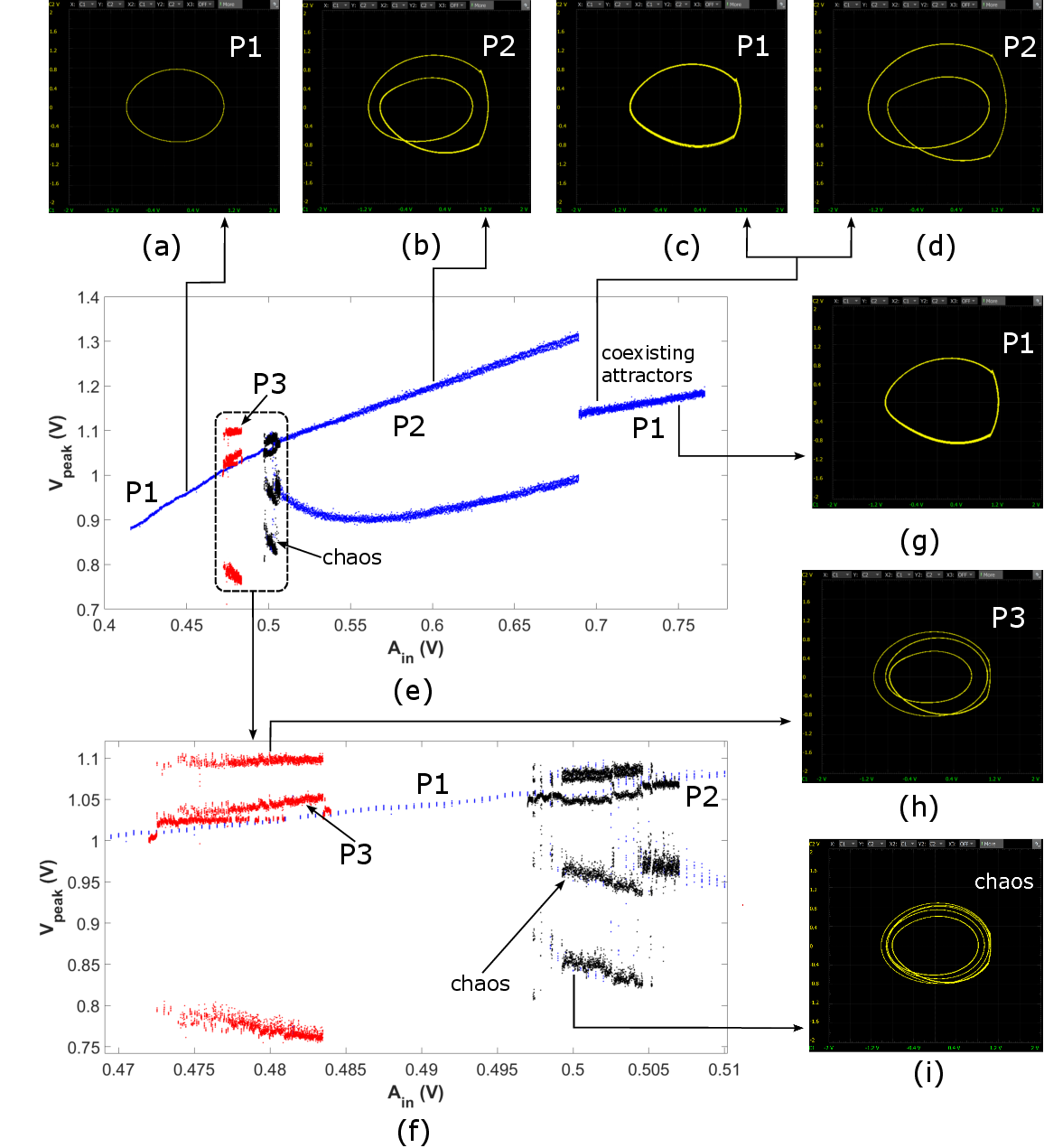}
	\caption{Experimental bifurcation diagram of the peak voltage \(V_{\text{peak}}\) vs. the input amplitude \(A_{\text{in}}\), shown for two distinct ranges: (e) \(0.416 \leq A_{\text{in}} \leq 0.766\) and (f) \(0.470 \leq A_{\text{in}} \leq 0.507\). Phase portraits, captured by a digital oscilloscope, corresponding to different values of \(A_{\text{in}}\): (a) P1T0 at \(A_{\text{in}} = 450\)~mV, (b) P2T1 at \(A_{\text{in}} = 600\)~mV, coexisting attractors at \(A_{\text{in}} = 700\) ~mV corresponding to (c) P1T1 and (d) P2T1, (g) P1T1 at \(A_{\text{in}} = 750\)~mV, (h) P3T1 at \(A_{\text{in}} = 480\)~mV, and (i) An aperiodic orbit at \(A_{\text{in}} = 500\)~mV.}
	\label{fig 18}
\end{figure}
Fig.~\ref{fig 18} presents the experimentally obtained bifurcation diagrams and the corresponding phase portraits. The bifurcation diagram in Fig.~\ref{fig 18}(e) consists of peak voltage values recorded for \(2103\) different input amplitudes within the range \(0.416 \leq A_{\text{in}} \leq 0.766\). A magnified bifurcation diagram highlighting finer variations in \(A_{\text{in}}\) over the intervals \(0.470 \leq A_{\text{in}} \leq 0.484\) and \(0.497 \leq A_{\text{in}} \leq 0.507\) is depicted in Fig.~\ref{fig 18}(f). These experimental results corroborate the presence of P1, P2, P3, and aperiodic orbits, which were numerically obtained in Fig.~\ref{fig 10}.  

The corresponding phase portraits illustrate various dynamical regimes:

- Fig.~\ref{fig 18}(a) represents a P1T0 orbit at \(A_{\text{in}} = 450\)~mV, characterized by before impact condition.

- Figs.~\ref{fig 18}(b) and \ref{fig 18}(g) depict P2T1 and P1T1 orbits at \(A_{\text{in}} = 600\)~mV and \(750\)~mV, respectively, each exhibiting a single impact.

- Figs.~\ref{fig 18}(c) and \ref{fig 18}(d) correspond to coexisting attractors P1T1 and P2T1 at \(A_{\text{in}} = 700\)~mV.

- Figs.~\ref{fig 18}(h) and \ref{fig 18}(i) illustrate a P3T1 orbit at \(A_{\text{in}} = 480\)~mV and a chaotic orbit at \(A_{\text{in}} = 500\)~mV.

\begin{figure}[h!]
	\centering
	\includegraphics[scale = 0.8]{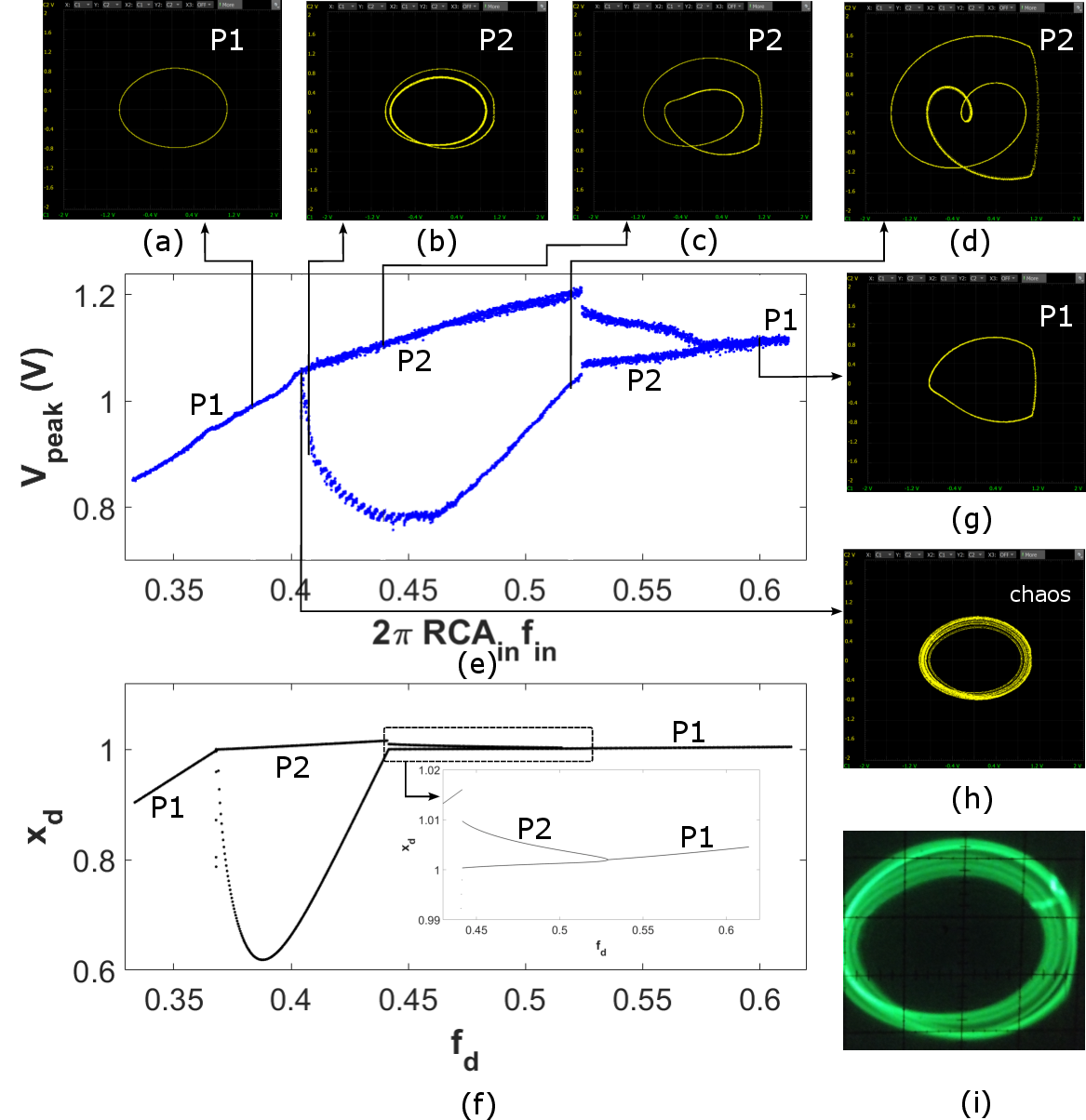}
	\caption{Bifurcation diagram of \( V_{\text{peak}} \) as a function of \( A_{\text{in}} \) obtained experientally in (e) \( 0.416 \leq A_{\text{in}} \leq 0.766 \) and numerically in (f). The corresponding phase portraits depict the following - (a) a P1T0 orbit at \( A_{\text{in}} = 480 \) mV, P2T1 orbits in (b), (c), (d) and (g) at $A_{\text{in}} = 495$mV, $A_{\text{in}} = 550$mV, $A_{\text{in}} = 650$mV and $A_{\text{in}} = 700$mV, (h) a P1T1 orbit at $A_{\text{in}} = 780$mV and (i) a chaotic orbit at $A_{\text{in}} = 486$mV.}
	\label{fig 19}
\end{figure}
Subsequently, the electronic circuit is configured to model the soft-impact oscillator with a high stiffness ratio of \(\beta = 50\) by taking $R_8 = 200\Omega$, as shown in Fig.~\ref{fig 19}. Fig.~\ref{fig 19}(e) is the bifurcation diagram consisting of peak voltage values recorded for \(701\) different input amplitudes within the range \(0.416 \leq A_{\text{in}} \leq 0.766\). Fig. \ref{fig 19}(f) is the numerically obtained bifurcation diagram provided for a direct comparison with the experimental results. The inset figure in Fig. \ref{fig 19}(f) shows the zoomed regions where P2 and P1 orbits are observed.

The corresponding phase portraits illustrate various dynamical regimes:

- Fig.~\ref{fig 19}(a) represents a P1T0 orbit at \(A_{\text{in}} = 480\)~mV, characterized by before impact condition.

- Figs.~\ref{fig 19}(b), (c) and (d) depict P2T1 orbits at \(A_{\text{in}} = 495\)~mV, $550$~mV and \(650\)~mV, respectively, each exhibiting a single impact.

- Fig.~\ref{fig 19}(g) correspond to a P1T1 orbit at \(A_{\text{in}} = 780\)~mV.

- Fig.~\ref{fig 19}(h) illustrate a chaotic orbit at \(A_{\text{in}} = 486\)~mV.

- Fig.~\ref{fig 19}(i) represents the experimentally obtained Poincar\'e section, demonstrating a finger-shaped attractor at an input amplitude of $A_{\text{in}} = 526$~mV.

\begin{figure}[tbh]
	\centering
	\includegraphics[scale = 0.8]{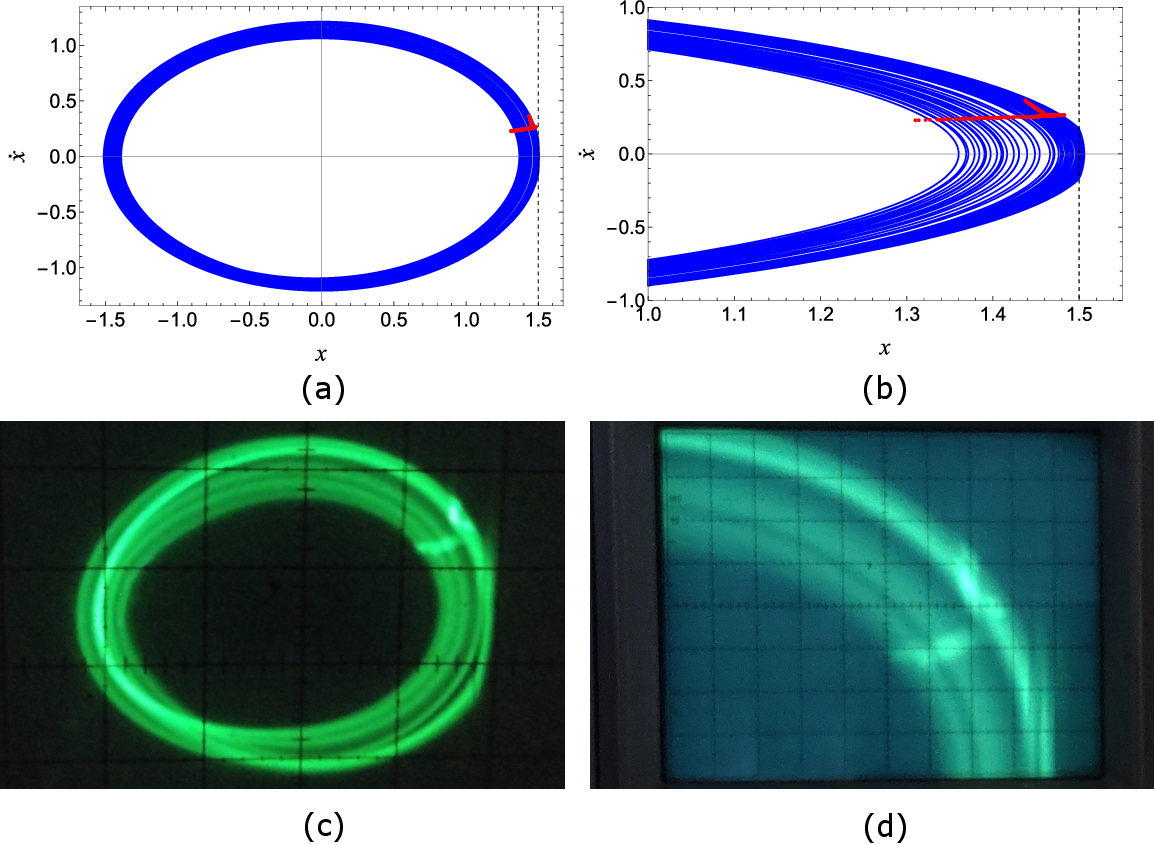}
	\caption{Poincar\'e sections for $\beta = 1.0$ obtained numerically for $f = 0.5513~\text{N}$ in (a) and (b), and experimentally for $A_{\text{in}} = 0.526~\text{mV}$ in (c) and (d).}
	\label{fig 20}
\end{figure}
Fig.~\ref{fig 20}(a) and Fig.~\ref{fig 20}(b) show the numerically obtained Poincar\'e section for $f = 0.55313$~N and stiffness ratio $\beta = 1$. These points were recorded stroboscopically at intervals of $2\pi/\omega$ and exhibit a finger-shaped attractor characteristic of near grazing dynamics in impact oscillators with the variation of stiffness from low to high values. Similar studies were conducted in~\cite{seth2020electronic}, except for cases where $\beta$ approaches unity. In Fig.~\ref{fig 20}(c) and Fig.~\ref{fig 20}(d), the Poincar\'e section is experimentally obtained using an analog oscilloscope, where the screen brightness is modulated via the $z$-axis input. The detailed methodology is provided in Appendix~\ref{app c}. The finger-shaped attractor is observed experimentally for the lower stiffness ratio $\beta = 1$ in Fig.~\ref{fig 20}(c) and Fig.~\ref{fig 20}(d).

The above-mentioned experimental results validate that the proposed inductorless circuit accurately replicates the vibro-impact dynamics of a soft-impact oscillator when the stiffness ratio (\(\beta\)) is unity. Similarly, for other \(\beta\) values, this circuit works perfectly. 

\section{Conclusion}
\label{sec 6}

In this paper, we have derived a closed-form expression for a higher-order transverse discontinuity mapping (TDM) in piecewise-smooth Filippov systems, incorporating a refined estimate of flight time. This improved estimate accurately predicts when perturbations near an impacting orbit reach the discontinuity boundary. The resulting quadratic expression reveals that only perturbations with a positive discriminant lead to impacts, offering a clear analytical condition for impact occurrence. In contrast, the first-order approximation yields real-valued flight times for all perturbations, failing to capture cases where perturbations do not reach the discontinuity boundary. Additionally, the higher-order TDM includes correction terms in the denominator that prevent divergence in low-velocity, near-grazing impacts. This leads to more accurate perturbation mappings, which are otherwise overestimated by the first-order saltation matrix.

A numerical scheme has also been developed to compute the higher-order saltation matrix, incorporating matrix inversions to account for higher-order terms. This results in an enhanced state transition matrix, which can be further extended to include corrections of order $\mathcal{O}(3)$ or higher. The corresponding monodromy matrix, evaluated along periodic orbits, facilitates stability analysis via its Floquet multipliers. For a schematic impact oscillator with a pre-stressed deformable barrier, this analysis successfully predicts instability thresholds consistent with both forward and backward bifurcation diagrams, capturing the emergence of period-doubling bifurcations and hysteresis-induced transitions. 

For aperiodic orbits, the Lyapunov spectrum, including the largest Lyapunov exponent, is estimated using the higher-order TDM without requiring an explicit analytical form of the map during impact. These numerical results show strong agreement with the underlying bifurcation structures, effectively identifying transitions to chaos as system parameters vary.

The theoretical framework is further validated through experimental implementation. A novel inductorless electronic circuit is designed to model a bilinear oscillator, emulating the dynamics of an impact oscillator interacting with a prestressed barrier. Unlike conventional LCR-based configurations, the proposed circuit employs operational amplifiers to integrate two first-order differential equations, allowing for accurate representation of discontinuity-induced bifurcations, even at low stiffness ratios. Moreover, the circuit successfully reproduces multiple coexisting stable attractors. Experimental observations, including phase portraits, Poincar\'e sections, and bifurcation diagrams, exhibit period-1, period-2, and period-3 dynamics under parametric modulation, showing strong agreement with numerical predictions.

\bibliographystyle{abbrv}
\bibliography{References}

\appendix
\section{Monodromy matrix}\label{app a}

\begin{algorithm}[H]
    \caption{Floquet multipliers from monodromy matrix $\mathbf{\phi}$} \label{algo 2}
    \begin{algorithmic}[!]
    \State 1. Initialize: $\mathbf{x}(0)$ ensuring $H(\mathbf{x}) \geq 0$ and $\omega$ or $\alpha$ \Comment{Bifurcation parameter}
    \State 2. Initialize: $\mathbf{\Phi} = \textit{I}_{n \times n}$, T$ = 100$, count $= 0$, $n_{max}$ \Comment{Maximum allowable impacts}
    \While{count $\leq$ $n_{max}$}
    \State Integrate: $\dot{\mathbf{x}} = \mathbf{F}(\mathbf{x})$
    \If{$H(\mathbf{x}) = 0$} \Comment{Occurrence of impact}
        \State Reset Map: $\mathbf{x} \gets \mathbf{R}(\mathbf{x})$
    \EndIf
    \If{$\dot{x} = 0$ $\&\&$ count $\geq n_{max}/2$}
        \State $T = \{t \in \mathbb{R}^1:x(t + \tilde{n}T) = x(t), \tilde{n}\in \mathbb{I} \}$ \Comment{Store $x$, $t$ for evaluation of time period $T$}
    \EndIf
    \If{count $=$ $n_{max}$ - 100} \Comment{Remove transients}
        \State Store: $\mathbf{x}_{init}$ $\gets \mathbf{x}(t)$ and $t_{init} \gets t$ 
    \EndIf
    \EndWhile
    \State Initialize: $\mathbf{x} \gets \mathbf{x}_{init}$ at $t_{init}$ \Comment{Begin at steady state} 
    \State Initialize: $r_0 \ll 1$, $\mathbf{Y}_1(0) = r_0 \textit{I}_{n \times n}$, $\mathbf{Y}_2(0) = \textit{I}_{n \times n}$ \Comment{Evaluation of $\mathbf{S}_2$ and $\mathbf{\Phi}_i$}
    \While{$t_{init} \leq t \leq (t_{init} + T)$} \Comment{Integrate over period $T$}
        \State Integrate: $\dot{\mathbf{x}} = \mathbf{F}(\mathbf{x})$, $\dot{\mathbf{Y}_1} = (\nabla \mathbf{F})^T\cdot\mathbf{Y}_1$ and $\dot{\mathbf{Y}_2} = (\nabla \mathbf{F})^T\cdot\mathbf{Y}_2$ 
        \If{$H(\mathbf{x}) = 0$} \Comment{Occurrence of impact}
        \State $\mathbf{\Phi}_i \gets \mathbf{Y}_2$ and $\mathbf{\Phi} \gets \mathbf{\Phi}_i\cdot\mathbf{\Phi}$
        \State Evaluate $\delta$ and $\mathbf{y}_+$ from $\mathbf{x}_4$ for all $\mathbf{y}_i$ \Comment{TDM}
        \State Evaluate $\mathbf{Y}_{+, impact}$ from $\mathbf{y}_+$ and $\mathbf{Y}_{-\text{, impact}} \gets \mathbf{Y}_1$
        \State Evaluate $\mathbf{S}_2 \gets \mathbf{Y}_{+, impact}\cdot\mathbf{Y}_{-, impact}^{-1}$ and set $\mathbf{\Phi} \gets \mathbf{S}_2 \cdot \mathbf{\Phi}$ \Comment{saltation matrix}
        \State Reset Map: $\mathbf{x} \gets \mathbf{R}(\mathbf{x})$
        \State Reinitialize: $\mathbf{Y}_1 \gets r_0 \textit{I}_{n \times n}$ and $\mathbf{Y}_2 \gets \textit{I}_{n \times n}$
        \EndIf
    \EndWhile
    \State 3. $\mathbf{\Phi} \gets \mathbf{Y}_2\cdot\mathbf{\Phi}$ \Comment{Evaluate monodromy matrix}
    \State 4. Evaluate eigenvalues of $\mathbf{\Phi}$ to get Floquet multipliers
   \end{algorithmic}
\end{algorithm}

\section{Lyapunov exponents}\label{app b}

\begin{algorithm}[H]
    \caption{Lyapunov exponent for hybrid systems using TDM} \label{algo 1}
    \begin{algorithmic}
    \State 1. Initialize: $\mathbf{x}(0)$ ensuring $H(\mathbf{x} \geq 0)$ and $\omega$ or $\alpha$ \Comment{Bifurcation parameter} 
    \State 2. Initialize: $\mathbf{y}_i(0)$ for $i \leq n$ using QRD 
    \State 3. Rescale: $\mathbf{y}_i(0) \gets r_0\times \mathbf{y}_i(0)$ and set $n_{max}$ \Comment{Maximum allowable impacts}
    \While{count $\leq n_{max}$}
    \State Integrate: $\dot{\mathbf{x}} = \mathbf{F}(\mathbf{x})$
    \If{$H(\mathbf{x}) = 0$} \Comment{Occurrence of impact}
        \State Evaluate: $\delta$, $\mathbf{R}(\mathbf{x})$ and $\mathbf{y}_+$
        \State Apply reset maps: $\mathbf{x} \gets \mathbf{R}(\mathbf{x})$ and $\mathbf{y}_i \gets \mathbf{y}_{i,+}$ \Comment{Implement TDM} 
    \EndIf
    \If{$\big( t \% \dfrac{2\pi}{\omega} \big) = 0$}
        \If{count $\geq n_{max}/2$}
            \State Store: $r_i \gets \dfrac{1}{r_0} \| \mathbf{y}_i \|$ \Comment{store the growth rate}
        \EndIf
    \State Reinitialize: $\mathbf{y}_i \gets r_0 \times \text{QRD of } \mathbf{y}_i$ 
    \EndIf
    \EndWhile
    \State 4: Evaluate: $\log_e r_i$ \Comment{Store all $log_e r_i$}
    \State 5: LE$_i \gets \dfrac{\omega}{2\pi} \times \text{Partial sum of } \log_e r_i$
    \State 6: $\lambda_i \gets \langle \text{LE}_i \rangle$ \Comment{Mean of all LEs}
    \end{algorithmic}
\end{algorithm}

\section{Circuit Implementation and Experimental Procedure}\label{app c}
\begin{figure}[tbh]
	\centering
	\includegraphics[scale = 0.7]{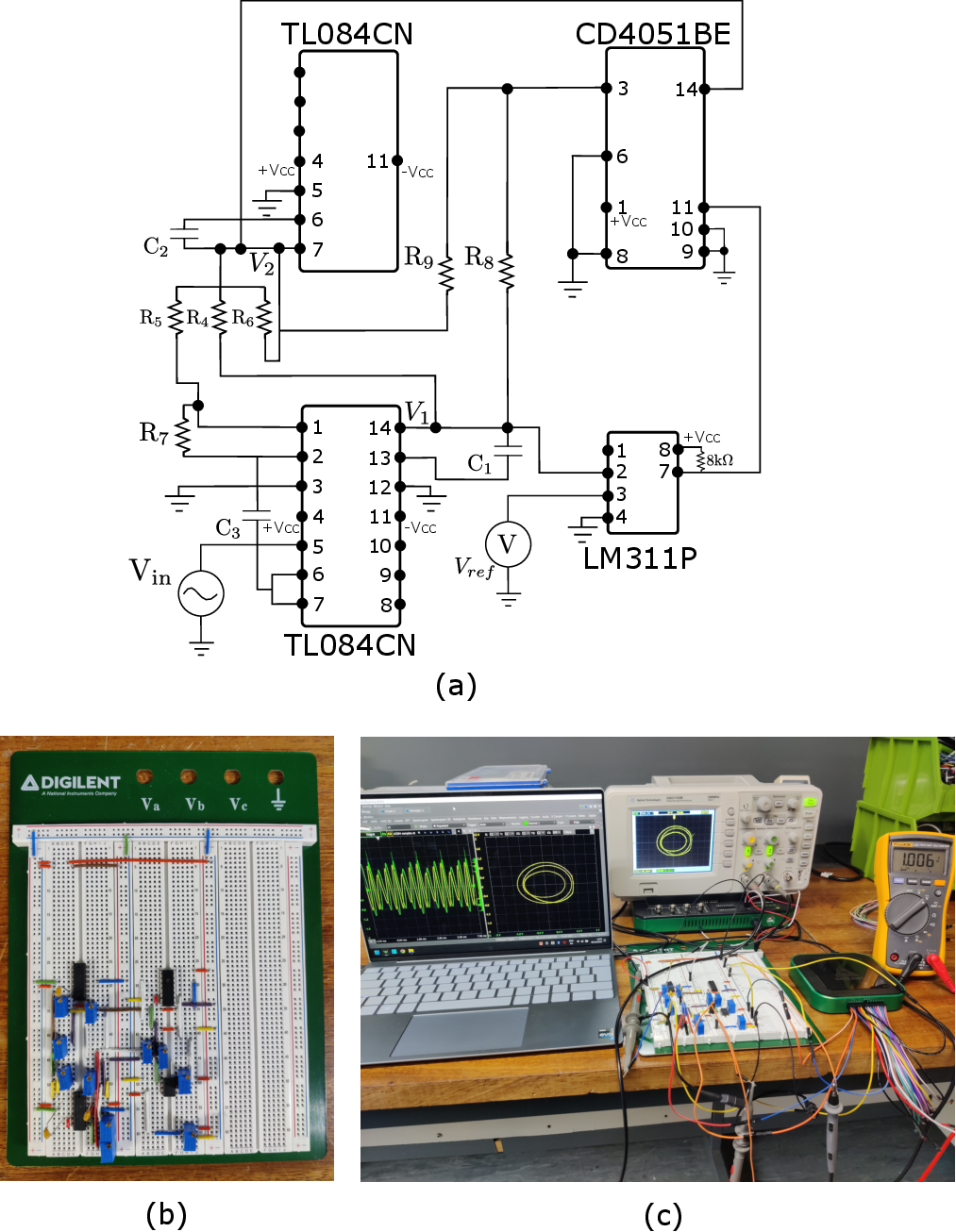}
	\caption{(a) The Circuit diagram showing the IC pin configurations. (b) Circuit implementation on the Digilent breadboard. (c) Experimental setup comprising the breadboard, a Digilent Discovery 3.0 data acquisition device, an oscilloscope, and a reference voltage of $1.006$~V.}
	\label{fig 21}
\end{figure}
Fig.~\ref{fig 21}(a) shows the circuit implementation of the design in Fig.~\ref{fig 16}(c), utilizing two TL084CN quad operational amplifier ICs. The discontinuity boundary is defined by the reference voltage $V_{\text{ref}}$, set to $1$~V. Switching in the bilinear oscillator is achieved using an LM311P comparator IC and a CD4051BE CMOS single 8-channel analog multiplexer/de-multiplexer with logic-level conversion. The implementation on a breadboard is depicted in Fig.~\ref{fig 21}(b). The complete experimental setup, including the breadboard, a Digilent Discovery 3.0 data acquisition device for signal generation and measurement, and a reference voltage set to $1.006$~V, is shown in Fig.~\ref{fig 21}(c).

The bifurcation diagram was experimentally obtained by interfacing the Digilent Discovery 3.0 data acquisition (DAQ) device with \textsc{MATLAB}. Each input signal $V_{\text{in}}$ is generated for a duration of one second and sampled at 200~Hz, with the corresponding peaks extracted in real-time. This efficient method enables real-time observation of discontinuity-induced bifurcation diagrams across a wide range of parameters while simultaneously allowing visualization of the phase portrait on an oscilloscope. A video demonstration of this procedure is provided as supplementary material.

The Poincar\'e points were experimentally captured using an analog oscilloscope equipped with $z$-axis modulation, which adjusts the brightness of the cathode-ray tube (CRT) display based on an external trigger signal. By applying an appropriate waveform to the $z$-axis input, one can modulate the brightness such that only specific segments of the trajectory are highlighted. In particular, a rectangular pulse waveform with suitable pulse width and frequency matching that of the input excitation $f_{\text{in}}$ was used to isolate and enhance the desired Poincar\'e sections. Synchronizing the pulse phase with that of the input excitation allows selective illumination of different portions of the continuous-time phase portrait, effectively yielding discrete-time Poincaré sections.

\end{document}